\documentclass[preprint2]{aastex}

\shorttitle{Speckle interferometry at Blanco and SOAR}
\shortauthors{Tokovinin, Mason, \& Hartkopf}

\begin{document}

\title{Speckle interferometry at the Blanco and SOAR telescopes in 2008 and 2009}

\author{Andrei Tokovinin}
\affil{Cerro Tololo Inter-American Observatory, Casilla 603, La Serena, Chile}
\email{atokovinin@ctio.noao.edu}

\author{Brian D. Mason\altaffilmark{1}, and William I. Hartkopf\altaffilmark{1} }
\affil{U.S. Naval Observatory, 3450 Massachusetts Avenue, NW,
 Washington, DC 20392-5420, USA}
\email{bdm@usno.navy.mil, wih@usno.navy.mil}

\altaffiltext{1}{Visiting Astronomer, Cerro Tololo Inter-American Observatory.
CTIO is operated by AURA, Inc.\ under contract to the National Science
Foundation.}

\begin{abstract}
The results of speckle interferometric measurements of binary and
multiple stars conducted in 2008 and 2009 at the Blanco and SOAR 4-m
telescopes in Chile are presented. A total of 1898 measurements of 1189 
resolved pairs or sub-systems and 394 observations of 285 un-resolved 
targets are listed. We resolved for the first time 48 new pairs, 21 of
which are new sub-systems in close visual multiple stars.
Typical internal measurement precision is 0.3\,mas in both 
coordinates, typical companion detection capability is $\Delta m \sim 4.2$ 
at 0\farcs15 separation. These data were obtained with a new 
electron-multiplication CCD camera; data processing is described in 
detail, including estimation of magnitude difference, observational
errors, detection limits, and analysis of artifacts. We comment on some newly 
discovered pairs and objects of special interest.
\end{abstract}


\section{Introduction}

Speckle  interferometry at  4-m telescopes  has provided  the  bulk of
binary star  measurements over the  last two decades,  giving material
for calculation of orbits  and other studies. Unfortunately, access to
4-m  telescopes  has been  intermittent,  especially  in the  southern
hemisphere  (speckle data from  the WIYN  telescope were  published by
\citet{Horch99,Horch02}). Here we present the results of two observing
programs carried  out in  2008-2009 to help  rectify this  problem. We
concentrate  on the  technique  and measurement  results, leaving  the
exploitation of these data  sets for further publications.  The general
neglect of  close southern hemisphere  binary stars has allowed  us to
determine new orbits for a  dozen pairs and correct preliminary orbits
for around one  hundred other pairs. These orbits  are currently being
prepared.


Continued measurements of visual binary stars are needed for many
reasons. One of the most obvious, yet most difficult tasks is to
establish the orbital elements of known binaries. Long orbital periods
where only a short arc is covered, on the one hand, and the lack of
continuous coverage of fast systems, on the other hand, prevent
calculation of orbits or cause erroneous orbits to be published.
Although stellar masses derived solely from visual orbits are
typically of inferior accuracy compared to other techniques, reliable
orbital elements are needed nevertheless in order to be able to
predict stellar positions and to study individual systems of
astrophysical importance (including those with planetary companions).
Multiplicity affects stellar evolution in many different ways, as
illustrated e.g. by the dramatic story of the quadruple system Regulus
\citep{Rappaport2009}.

Modern hydrodynamical simulations open the way to understanding the 
formation of binary and multiple systems \citep[e.g.][]{Bate08}, so good 
observational data on multiplicity statistics become critical for further
progress.
The current census of stellar multiplicity is incomplete even in the
solar neighborhood. 
In updating the seminal work of \citet{DM91}, \citet{Raghavan09}
determined that the fraction of triple and higher-order multiples
among G-type dwarfs within 25\,pc of the Sun was underestimated by
as much as two times. These additional triples and quadruples were
found around systems which were previously considered binary; i.e.,
the number of singles remained essentially fixed. For a well studied
group, the best location to find new companions is around double or
higher order multiple systems and not single stars. Therefore, we
also searched for new close sub-systems in wide visual binaries,
focusing on binaries in the solar vicinity.

Relative orientation of the orbits in triple
stars is an indicator of their formation mechanisms and dynamical
evolution \citep{ST02,FT07}. One of the goals of the present program
was to increase the number of multiple stars with known sense of
relative orbital motion in sub-systems by getting additional
observations. 

In any statistical study, it is important not only to detect new
companions, but also to establish the detection limits, so that an
absence of companions can be translated into constraints on their
parameters. The linearity of the detector allows us to here establish
reliable detection limits for each observation, through careful data
analysis and modeling. Owing to the new detector and data processing,
our observations reach larger magnitude difference than previous
speckle measurements. We also provide relative photometry of the
components.

Speckle interferometry at the Southern Astrophysical Research (SOAR)
telescope was started in 2007 with tests of a new high-resolution
camera \citep{TC08}. In the near future, this camera will work jointly
with the adaptive-optics system to reach diffraction-limited
resolution on faint targets in the visible. Meanwhile, it was used as
a stand-alone instrument.

Observing  time for  the USNO  intensified CCD  speckle camera  at the
Blanco  4-m telescope  at CTIO  was allocated  in July  2008  to cover
several programs,  ranging from orbit calculation  and improvement, to
observation of  nearby faint red  and white dwarfs and  subdwarfs, and
also   to  the   completion  of   a   speckle  survey   of  nearby   G
dwarfs. However, owing to  new export regulations, the equipment could
not be sent  to CTIO in time  for the run, despite all  efforts of the
NOAO  administration. The new  camera had  to be  used instead.  For a
number of reasons,  it was not possible to  reach the faintest targets
with  this new  camera, so  the program  had to  be modified  ``on the
fly". The  nearby, faint targets had  to be abandoned  and to preserve
uniformity  of the  sample,  most of  the  nearby G  dwarfs were  also
dropped.  Further,  failure   of  the  Blanco  atmospheric  dispersion
corrector (ADC) restricted observing  to a smaller than desired region
around the zenith. Nevertheless, a large number of useful measurements
were obtained.

\begin{figure*}
\epsscale{2.0}
\plotone{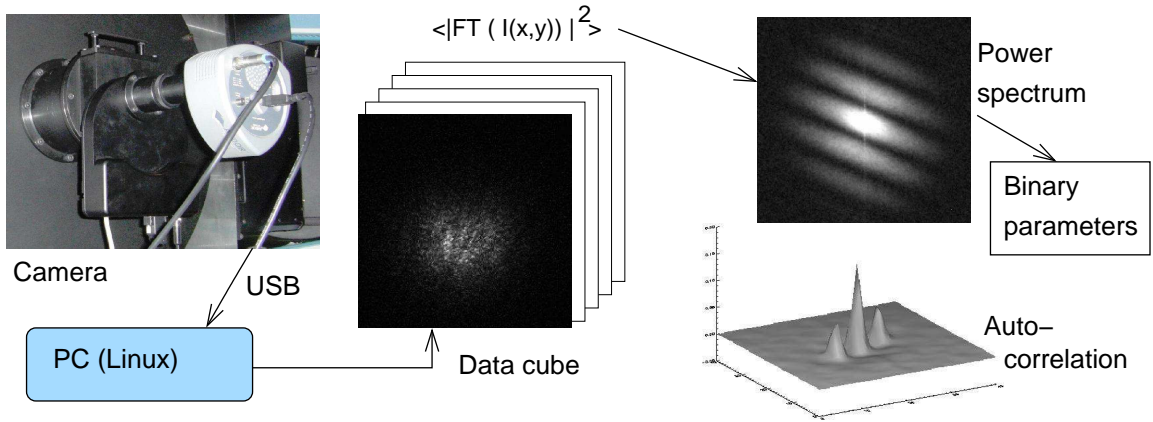}
\caption{Schematic representation of data acquisition and processing.
\label{fig:flow}}
\end{figure*}

We present the observational technique in Sect.~2, starting with the
instrument description and then detailing various data processing
steps (Fig.~\ref{fig:flow}). A new method of establishing the
detection limits is described and some effects which can lead to false
detections are studied. The main results are presented in Sect.~3,
including comments on some objects of interest and new discoveries.

\section{Observations and data analysis}

\subsection{Speckle camera}

The observations reported in this paper were obtained with the
{\it high-resolution camera} (HRCam) -- a fast imager designed to work at
the SOAR telescope, either with the SOAR Adaptive Module or as a
stand-alone instrument. The HRCam is described by \citet{TC08}; here
we recall its main features.

HRCam uses a CCD detector with internal electron multiplication -- an
EMCCD. The {\it Luca} camera from
Andor\footnote{\url{http://www.andor.com}} was chosen for its low
cost, fast frame rate, and simple signal interface via a USB port.
The CCD has 658x496 10-micron pixels. It is cooled thermoelectrically
to $-20^\circ$C, resulting in a very low dark current for the short
exposure time used here (except in a small number of {\it hot
pixels}). We used an EM gain of 44, so the readout noise of 14
electrons is effectively reduced to 0.3\,el. The quantum efficiency of
this detector is around 0.5.

HRCam consists of the detector, mechanical structure, filter wheel, 
and optics. The $f/16$ beam coming from SOAR is collimated by a 50-mm
negative achromat (Barlow lens) and refocused by a 100-mm positive
lens, doubling the effective telescope focal length and providing a
pixel scale of 15\,mas. At the Blanco $f/8$ telescope, we replaced
the negative lens by a positive achromat with 25-mm focal length to get
adequate sampling. 

HRCam has no corrector for atmospheric dispersion. Most observations were 
obtained through a Str\"omgren $y$ filter, while brighter stars were also 
observed through an H$\alpha$ interference filter, especially at large 
zenith distances. A few observations were taken through standard wide-band 
$V$, $R$ or $I$ filters.  Table~\ref{tab:filters} lists the central
wavelength and bandpass of each filter as measured (except for the
$I$-filter where the bandpass is determined by the detector cutoff in
the red)\footnote{The filter transmision curves are given in the
  instrument Manual,
  http://www.ctio.noao.edu/new/Telescopes/SOAR/Instru\-ments/SAM/archive/hrcaminst.pdf}.

\begin{table}
\center
\caption{Filter information}
\label{tab:filters}
\medskip
\begin{tabular}{c c c} 
\tableline\tableline
Filter    & Central Wavelength & FWHM,  \\
          & (nm)               & (nm)   \\
\tableline
$V$         & 517.2              & \phn84.2\phn \\
$y$         & 550.7              & \phn21.6\phn \\
$R$         & 596.1              &    121.2\phn \\
H$\alpha$ & 657.3              & \phn5.04 \phn \\
$I$         & 774.4              &  ---         \\
\tableline
\end{tabular}
\end{table}

The camera software developed by R.~Cantarutti works on a PC computer under the
Linux operating system. All basic functionality is provided, including the filter
wheel and detector control and connection to the telescope control
system. We used the detector in free-run continuous
mode. The central region of 200$\times$200 pixels may be read out with
exposure time of 20\,ms at a rate of 31.8\,ms per frame, or a
larger 400$\times$400 region may be read at a 43.1\,ms frame rate. The
required number (typically 400) of frames is grabbed during $13 - 17$
seconds and written to the disk. Immediately after the acquisition, a
quick-look power spectrum is calculated and displayed, showing the
quality of the data and the fringes for resolved binaries.

\subsection{Observing procedure}

Observing runs are listed in Table~\ref{tab:runs}. As noted above, the
use of HRCam at the Blanco telescope was not foreseen. The software
was not interfaced to the telescope, so information missing in the
FITS headers (acquisition time, object coordinates, zenith distance)
was retrieved later using our logbook. A few object identification
errors were corrected {\it post factum}, but some may still remain
undetected. We planned to use the ADC
of the Blanco telescope, but found it to be nonoperational, so the
dispersion remained uncorrected and the program was restricted to smaller
zenith distances. All 5 nights of the Blanco run were
clear, with variable seeing.

\begin{table}
\center
\caption{Observing runs}
\label{tab:runs}
\medskip
\begin{tabular}{l l c r} 
\tableline\tableline
\multicolumn{1}{c}{Run}      & \multicolumn{1}{c}{Dates}             & \multicolumn{1}{c}{Pixel Scale} & $N_{dat}$ \\
         &                   & (mas)  &           \\
\tableline
Blanco08 &  14-18 Jul 2008  & 16.35  & 2131      \\
SOAR08a  &  8 Aug 2008    & 15.23  &  328      \\
SOAR08b  &  6-8 Oct 2008 & 15.23  &  985      \\
SOAR09   &  4-6 Apr 2009   & 15.23  & 1416      \\
\tableline
\end{tabular}
\end{table}

The SOAR telescope is located at the Cerro Pach\'on mountain. The
shape of its thin 4.1-m mirror is controlled actively using a bright
star at the beginning of the night and using look-up tables for the
rest of the night. HRCam was installed at the Nasmyth focus and
received the light after 3 reflections. SOAR has an alt-azimuth mount;
field rotation at the Nasmyth focus is compensated by
counter-rotating the instrument. The position angle of the rotator is
recorded in the headers for subsequent calculation of angles
on the sky.

In preparation for the SOAR runs, we tested the system during a technical
night on August 8, 2008. Some useful data were collected during this
night, mostly on binaries with known orbits. During the two scheduled
3-night runs in October 2008 and April 2009, the sky was clear, the
wind speed was low, and the seeing was generally good, with the width
of the best long-exposure images as small as 0$''$\llap.70.

Efficient use of the allocated telescope time required good
preparation of the observing program and a substantial effort on the
part of the telescope operators. For example, during the April run, we
observed 551 objects on 3 nights, spending less than 3.5 minutes per
star on the average. Each observation included telescope slew, object
acquisition, data recording, and quick-look analysis. We recorded at
least two data cubes of each object and analyzed each cube
independently. This strategy helped to verify new companion detections
and to estimate observational errors. The last column of
Table~\ref{tab:runs} lists the total number $N_{dat}$ of data cubes taken during
each run.

The faintest stars observed were around $V = 10^m$, with a few
exceptions. For example, the components of DON~93 BC, $V=12^m$ and
$13.1^m$ at 0$''$\llap.76 separation, were measured reliably with an
accuracy of 1\,mas under good seeing. Companions of $V\sim13^m$ were
measured routinely if the primary star was brighter than
$9^m$. Paradoxically, the presence of a bright primary increases the
sensitivity for the secondary because the speckle signal is
proportional to the product of photon fluxes from both companions.
For faint stars, sensitivity could be gained at the expense of
resolution by increasing the exposure time (up to 100\,ms) and by
observing in wide-band filters. This last resource was, unfortunately,
not available to us because it requires a working compensator for
atmospheric dispersion.

\subsection{Calculation and modeling of the power spectra}

Each data cube typically contains $K=400$ images of 200$\times$200
or 400$\times$400 format, with 14 bits per pixel. The average dark
frame (with the same exposure time), called {\it bias}, is subtracted
from each image to remove the fixed offset of $\sim$505
analog-to-digital units (ADU) and the dark current. The average bias
is stable to within 1~ADU. Pixels in the bias frame which are more
than 10 ADU above the average level are identified as ``hot''; these
pixels in the images are replaced by the average of neighboring
pixels. Finally, all pixels below 17 ADU (twice the readout noise) are
set to zero in order to reduce the influence of remaining pattern and
readout noise. The optimum threshold value was selected after some
trials and then applied to all data. Single photons produce signals
well above this threshold. Despite the thresholding and the very low
dark current, each 200$\times$200 frame contains some 40 
photon events at random locations, apparently created in the
electron-multiplication process. This additional background is the
main factor which prevents observing very faint stars with long
accumulation times. For a subset of bright stars observed without EM
gain, only a fixed bias of 505~ADU was subtracted, without any
thresholding.

\begin{figure}
\epsscale{1.0}
\plotone{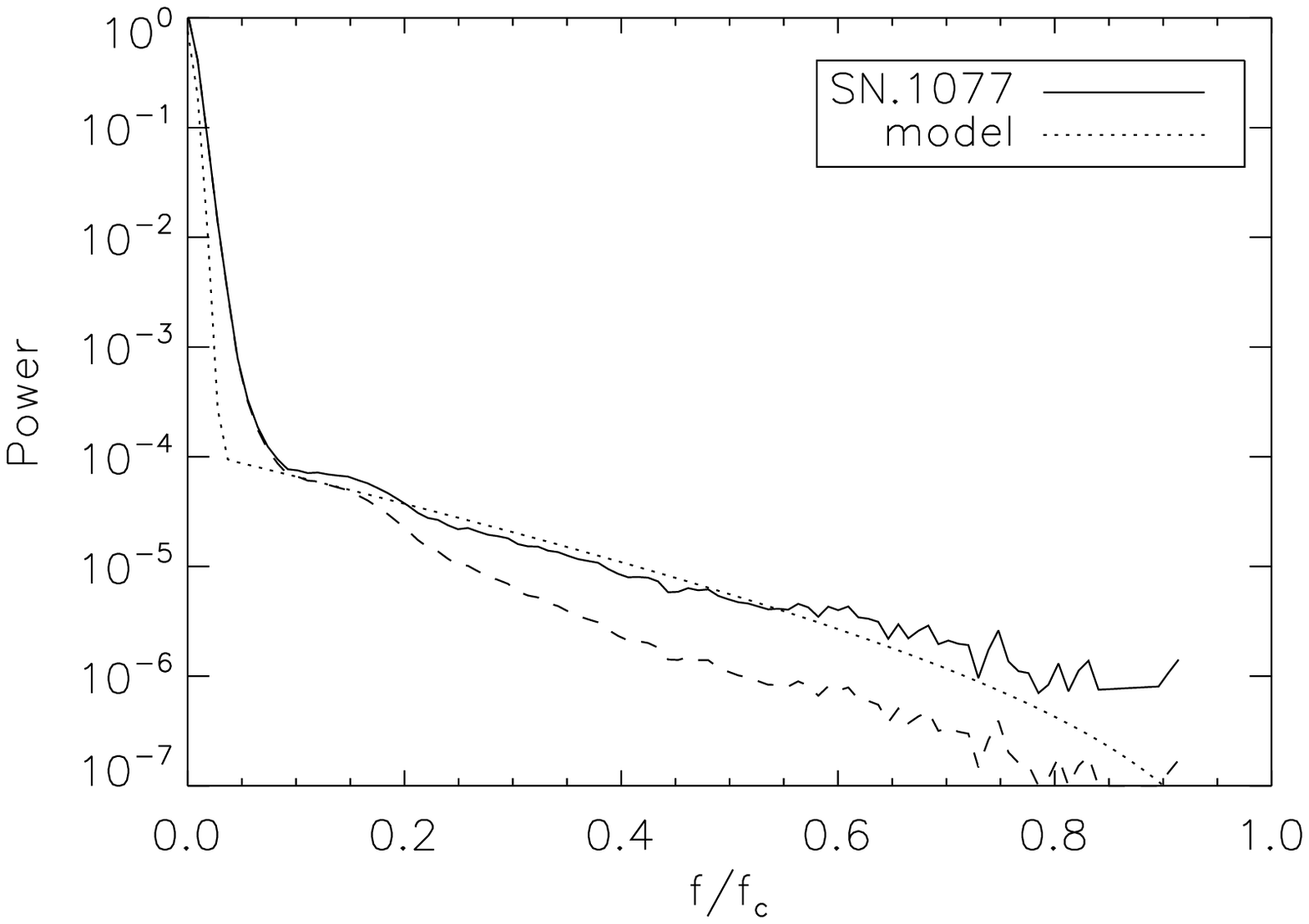}
\caption{Example of a power spectrum. Azimuthally averaged $P(f)$ is
plotted in dashed line (raw) and in full line after division by
$\overline{T}_{AD}( f)$. 
The model (\ref{eq:mod}) is plotted in dotted
line. Observations at SOAR in the $y$ filter, zenith distance $z=50^{\circ}$\llap.7. 
\label{fig:pow}}
\end{figure}

The power spectrum (PS) of an image
cube is calculated by summing the square modulus of the Fourier
Transform of each image,
\begin{equation}
P(f_x,f_y) = \frac{C}{K}\; \sum_{i=1}^{K}| \tilde{I}_i(f_x,f_y) |^2 ~.
\label{eq:pow}
\end{equation}
The spatial frequencies $f_x, f_y$ correspond to the elements of
square discrete arrays. The normalization constant, $C$, is determined
from the condition $P(0,0)=1$. In the following, it is convenient to use
normalized frequencies $\kappa = f/f_c$, where $f_c =
D/\lambda$ is the cutoff frequency, $D$ is the telescope diameter and $\lambda$ is the 
central wavelength of the filter passband. 

It is well known that the PS of a single bright star $P_0(f_x,f_y)$
contains two components: a strong signal at low spatial frequencies $f
< r_0/\lambda$ which corresponds to the seeing-limited image ($r_0$ is
the Fried parameter) and a high-frequency component extending up to
$f_c$ and produced by the speckle structure. An additive component
$P_{noise}$ is produced by the photon noise. It is easily estimated by
averaging the PS values at $f > f_c$. This additive term is then
subtracted from the PS.

Knowledge of the {\it speckle transfer function} (STF) $P_0$ is needed
for fitting a binary-star model to the data (see below). Single
reference stars are sometimes observed for this purpose. However, the
STF is not stable in time and depends on such factors as seeing
conditions, atmospheric dispersion (AD), telescope aberrations,
vibrations, etc. For each PS, we describe the STF in the
high-frequency zone by an empirical model, obviating the need for a
reference star and accounting for the changing conditions
automatically. The bias term $P_{noise}$ is subtracted from
$P(f_x,f_y)$, then the PS is averaged azimuthally, leading to the
one-dimensional function $\overline{P}(\kappa)$. A very simple
2-parameter model
\begin{equation}
\log_{10} P_r(\kappa) \approx \log_{10} [\overline{T}_{AD}(\kappa) T_0(\kappa)]  + p_0 + p_1 \kappa 
\label{eq:mod}
\end{equation}
is fitted in the range $\kappa_{\rm min} < \kappa < \kappa_{\rm max}$.
Typically, we select $\kappa_{\rm min} = 0.2$ and $\kappa_{\rm max} = 0.8$,
but for the noisy data the upper limit is reduced. Here $T_0(\kappa)
= 2/\pi[ \arccos \kappa - \kappa \sqrt{ 1 - \kappa^2 } ]$ is the
diffraction-limited transfer function of an ideal telescope (the
central obstruction is ignored), and $ \overline{T}_{AD}(\kappa)$ is the
azimuthally-averaged deterministic blur $T_{AD}(f_x,f_y)$ caused by
the atmospheric dispersion (AD), 
\begin{equation}
T_{AD}({\bf f}) \approx \exp [ - 2 \pi^2 ( {\bf f} {\bf x}
 /2.506)^2 ]. 
\label{eq:AD}
\end{equation}
The AD blur is represented by a Gaussian function. The length of the
blur vector $|{\bf x}| = [n(\lambda_1) - n(\lambda_2)]/p \; \tan z $ (in
pixels) is known for the zenith distance $z$, refractive index of air
$n(\lambda)$, filter bandwidth limits $\lambda_1$ and $\lambda_2$, and
the pixel size $p$. The direction of the vector ${\bf x}$ is known
from the calculated parallactic angle and the detector orientation.
Figure~\ref{fig:pow} illustrates power-spectrum modeling.

The parameter $p_0$ shows the level of the high-frequency component of
the PS extrapolated to zero frequency. The theoretical PS model
predicts that $(D/r_0)^{-2} = 0.435\, 10^{p_0}$, leading to an
estimate of the seeing conditions relevant to each data cube from the
$p_0$ values. These seeing estimates match quite well the half-width
of the re-centered long-exposure images calculated from the data cubes.
The second parameter, $p_1$, shows how fast the high-frequency
component of the PS is decreasing. For an ideal speckle pattern $p_1
= 0$, but in reality finite exposure time, finite bandwidth and other
factors lead to $p_1 < 0$.

The {\it synthetic} STF is thus calculated as 
\begin{equation}
P_{0, syn}({\bf f}) = T_{AD}({\bf f}) T_0(f) 10^{p_0 + p_1
 (f/f_c)}
\label{eq:P0syn}
\end{equation}
for the selected frequency range and $f = | {\bf f}|$. Alternatively,
we use the azimuthally-averaged observed PS $\overline{P}( f)$ as a
reference, with an additional multiplier $T_{AD}({\bf f})/
\overline{T}_{AD}( f) $ to account for the AD. For binaries with
separations above 0\farcs1 the radially-averaged reference is normally
chosen, while synthetic reference is used for closer binaries.


\subsection{Fitting parameters of binary and triple stars}

\begin{figure}
\epsscale{1.0}
\plotone{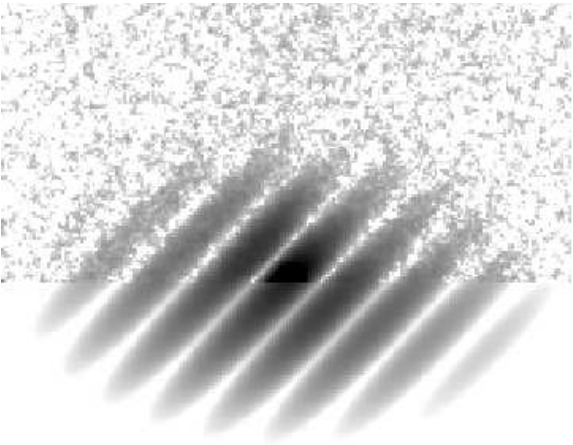}
\caption{The PS of binary star FOX~102AB ($\rho$ = 0$''$\llap.17, $\Delta m
 = 0.45$) is displayed, replaced by the fitted model (\ref{eq:bin})
 with synthetic STF in the lower half. The intensity scaling is
 inverse logarithmic from $10^{-6}$ to $10^{-3}$. The AD blur in the
 $y$-filter was 3.4 pixels ($ z = 38^\circ$) in approximately vertical
 direction, causing elongation of the PS.
 \label{fig:example}}
\end{figure}

The PS of a binary star shows characteristic fringes. It is more
practical, nevertheless, to detect companions in the auto-correlation
functions (ACFs) calculated from the PS by Fourier transform. The
two-component structure of the PS is carried to the ACF which
consists of a wide seeing pedestal and three narrow peaks (in the case
of binary star). The pedestal can be removed by setting to zero the PS
at low spatial frequencies, e.g. at $f < 0.2 f_c$. Such crude
filtering leads to ``ringing'' in the ACF. To avoid it, we divide the
PS by its azimuthal average $\overline{P}(f)$ at low frequencies where
it exceeds the extrapolated level of the speckle signal, $10^{p_0}$,
and apply additional Gaussian damping to further reduce the low
frequencies. The {\it filtered} ACFs are then computed from the
filtered PS and are used together with the PSs for binary-star
analysis. 

The parameters of a binary star are the time of observation, $T$,
separation, $\rho$, position angle, $\theta$, and magnitude
difference, $\Delta m$. The first number is arbitrarily precise, the
next two numbers are combined in a 2-dimensional vector ${\bf r} =
(\rho \cos \theta, \rho \sin \theta)$. The observed power spectrum
$P({\bf f})$ (after subtraction of $P_{noise}$) is fitted by a model

\begin{equation}
P_{mod}({\bf f}) = P_0({\bf f}) \; [ A + B \cos (2 \pi {\bf f} {\bf r}) ] ~, 
\label{eq:bin}
\end{equation}
where $ P_0({\bf f})$ is the STF and the coefficients $A$ and $B$ are
related to the magnitude difference. The position angle is determined
only modulo $180^\circ$, so a change of quadrant is always possible. 

Fitting of the model (\ref{eq:bin}) to the PS is done by the
Levenberg-Marquardt method in the frequency range $\kappa_{min} <
\kappa < \kappa_{max}$ over the upper half-plane $\kappa_y \ge 0$, due
to the symmetry. The error of the measured PS at each point $j$,
$\sigma^2_j = (P_j + P_{noise})/K$ is taken into account ($K$ --
number of images in the data cube). The quality of the fit is
evaluated by the normalized sum of residuals $\chi^2/N = (1/N) \;
\sum_j (P_j - P_{mod, j})^2/ \sigma_j^2$, $N$ being the total number
of the fitted points in the frequency plane. For noisy data, we obtain
$\chi^2/N \sim 1$, but for bright stars the un-modeled systematic features
of the STF (e.g. details caused by telescope aberrations) dominate the
residuals, leading to $\chi^2/N$ values up to
20. Figure~\ref{fig:example} shows the example of a binary-star PS and
its fitted model. The fact that AD is explicitly included in the 
model helps to distinguish true close companions from the elongation
of speckles produced by the AD.

A model of a resolved triple star is fitted in a similar way, but the
number of parameters is larger. Initial values of the fitted
parameters are determined by clicking on the companion(s) on the
displayed ACF. In the data table we list positions and magnitude
differences of triple-star pairs relative to the brightest component,
not the photo-centers of close sub-systems. For example, the
measurements of J01198$-$0031 = STF~113~A,BC at 1\farcs67 in fact refer
to the pairing (A,B), not to the center of the inner sub-system
FIN~337~BC. Accordingly, in the data tables we designate the wide pair as STF~113~AB. 
%
The quadrants in a triple system can be changed jointly,
but not individually. When the quadrant of the slowly moving outer
pair is known from previous observations, the quadrant of the inner
pair can be established without ambiguity.

The fitting  program provides estimates  of the parameter  errors. Yet
another estimate comes from  the variance of $N$ measurements obtained
on the same night in  the same filter, $\sigma_x^2 = (N-2)^{-1} \sum_N
(x - \overline{x})^2$.  Mostly, $N=2$  (two data cubes).  We adopt the
larger of these  two errors and list them in the  data table. For 1846
measurements with $N \ge 2$,  the median errors of companion positions
are 0.3\,mas in  both radial and tangential directions,  while 75\% of
the errors are smaller than  0.8\,mas. For a subset of 49 measurements
with $N \ge 4$ where the  estimates of the variance are more reliable,
the median errors in tangential and radial directions are 0.5\,mas and
0.6\,mas,  respectively.   The error  in  separation exceeds  7.5\,mas
(half-pixel) in only 16 cases where the companions are either close or
very faint. The listed errors are  {\em internal}, they do not take into
account calibration uncertainties or other systematic effects.  During
the Blanco  run, some bright  pairs were observed repeatedly,  and the
argeement  between these  measurements  is quite  good.  For  example,
A~417 (WDS  23052$-$0742) shows the scatter  $\sigma_\rho = 0.30$\,mas
and $\rho \sigma_\theta = 0.23$\,mas from 6 measurements over 4 nights
in two filters.

\subsection{Calibration}

\begin{figure}
\epsscale{1.0}
\plotone{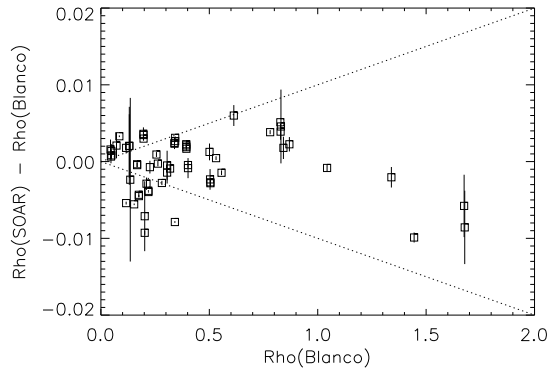}
\caption{Inter-comparison of separations measured in the Blanco08 and
 SOAR08b runs on 107 common pairs. The dotted lines show $\pm 1$\%
 deviations in the pixel scale. The estimated measurement errors are
 shown by the vertical lines.
\label{fig:BS08}}
\end{figure}

Accurate  knowledge of  the detector  pixel scale  and  orientation is
needed to convert binary-star  parameters from fitted values in pixels
to absolute positions  on the sky. Speckle measurements  at the Mayall
4-m  telescope at Kitt  Peak National  Observatory were  calibrated by
means of  a double-slit mask  \citep{Hrt2000a}, while speckle  data at
the Blanco  telescope were traditionally  tied to this  calibration by
observing  common  pairs.  Originally,  we  intended  to observe  many
binaries with known  orbits to calibrate our runs.  However, it turned
out  that the  quality  of the  available  binary star  orbits is  not
adequate.  By calibrating  against  orbits, we  relate modern  precise
measurements to the  historical data of much lower  accuracy which may
also contain systematic errors.

A comparison between the Blanco08 and SOAR08b runs has revealed a
disagreement of the pixel scale at the 3.5\% level, despite independent
calibration of each run with $\sim 100$ orbits. In the face of this
discrepancy, we calibrated the SOAR09 run by projecting into the
telescope a fringe pattern formed by two coherent point sources
attached to the telescope spider. The baseline of this interferometer
$b = 0.4999\,$m was accurately measured, the wavelength of the green
laser $\lambda = 532.2$\,nm is known, so the fringe period $\lambda/b
= $0\farcs2196 is known as well. About 30 fringes fit into the
400$\times$400 pixel field. The position of the fringe peak in the PS
of these data cubes is found by a simple centroid, leading to the
determination of the pixel scale and detector orientation.

The results for each of the 8 series of fringe measurements are very
consistent internally, but do show a spread between the series
amounting to 0.5\% in scale and $0\fdg2$ in angle. We attribute these
differences to small imperfections in the fringe pattern caused by
optical defects and aberrations in the beam path. The pixel scale of
15.23\,mas is finally adopted for all SOAR runs. We measured also the
effective pixel size of the HRCam detector (through its optics) by
illuminating the device with a laser and determining the angle between
the beams diffracted back by the pixel grid. The nominal
projected pixel size of $5.00 \pm 0.025$\,$\mu$m was confirmed. Using
the effective focal length of the SOAR known from its optical prescription,
$F=67.834$\,m, we obtain the pixel scale of 15.20\,mas in agreement
with the laser calibration.

The detector orientation in the SOAR09 run was independently checked
by observing stars at large zenith distances with wide filters. The
resulting PS is elongated perpendicularly to the AD direction which
itself is known. The elongation angle is determined by correlating
the observed and modeled PSs in a certain range of spatial frequencies
and finding the angle where the correlation reaches maximum. It
turned out that most consistent results are obtained by considering
the mid-range spatial frequencies between 0.1 and 0.3$f_c$, while at
higher frequencies the direction of the elongation is possibly
affected by telescope vibrations (see below). The resulting detector
angle determined from AD is $1\fdg32 \pm 0\fdg1$, to be compared
to $1\fdg37 \pm 0\fdg1$ measured from fringes and $1\fdg52 \pm
0\fdg2$ from orbits. In this case, all three methods agree very
well. 

\begin{figure}
\epsscale{0.9}
\plotone{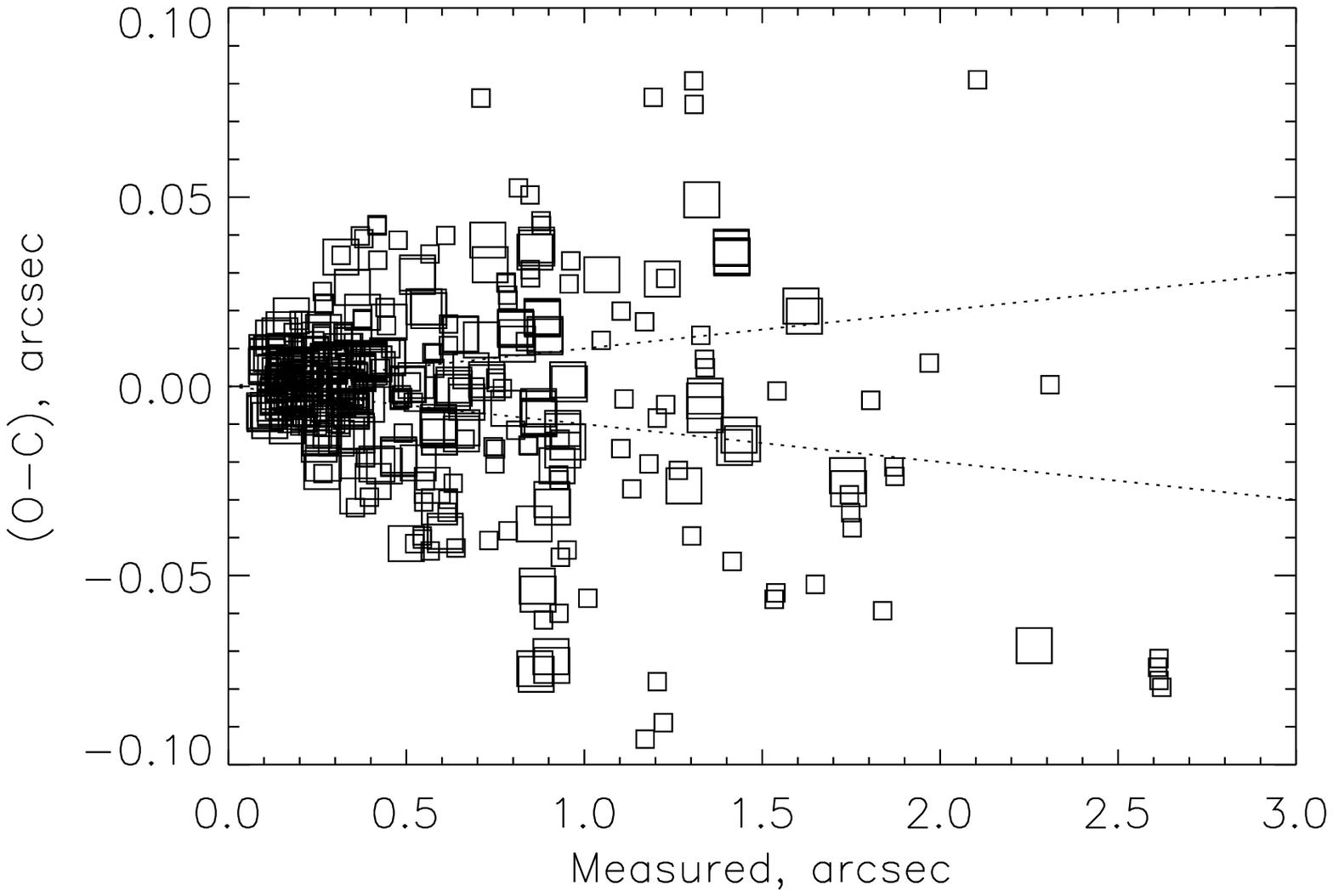}
\plotone{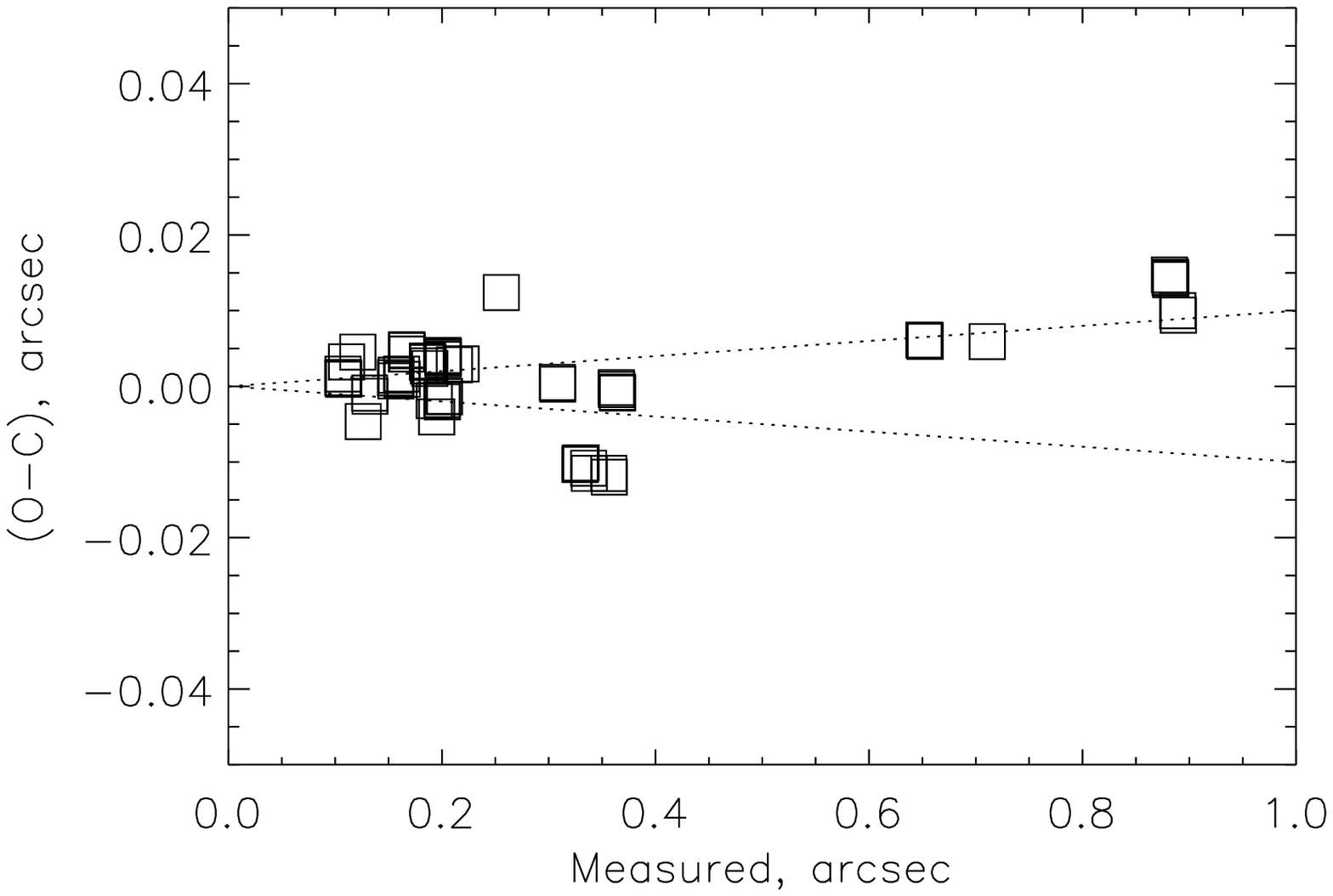}
\caption{Residuals (O$-$C)  in separation  with respect to  orbits. In
  the upper  panel, the whole data  set (432 points)  is plotted, with
  orbits  of grades 3  and higher  as larger  squares. The  two dotted
  lines  indicate $\pm 1$\%  range in  the pixel  scale. In  the lower
  panel, only the 64 points with orbits of grade 1 are retained.
 \label{fig:rho}}
\end{figure}

The  pixel scale of  the Blanco  run was  adjusted using  107 binaries
measured also  in October 2008  (Fig.~\ref{fig:BS08}).  The unweighted
rms difference $\rho_{Blanco} -  \rho_{SOAR}$ after adjustment is only
3.5\,mas. A similar level of  agreement is seen between other pairs of
runs.   This is  the upper  limit for  the {\em  external} measurement
errors (part  of the  difference is caused  by the motion  of binaries
between the runs).  In contrast, the residuals in $\rho$ to the orbits
(Fig.~\ref{fig:rho}) show a  large scatter for the whole  data set and
for  the  individual  runs,  precluding  accurate  scale  calibration.
Calibration only  with the orbits of  grade 1 does not  help.  The rms
scatter of  the O$-$C residuals  in $\rho$ in  Fig.~\ref{fig:BS08}b is
still 8.3\,mas for 32 points  with $\rho < 0\farcs4$, much larger than
the  difference between  the  runs.  The residuals  to  orbits in  the
tangential direction are around 5\,mas.

The detector orientation changes  slightly at each installation of the
camera,  so it  has to  be calibrated  for each  run.  Considering the
comparisons  with  orbits, the  difference  in  $\theta$ between  wide
binaries measured  commonly in pairs  of runs, and the  secure angular
offset  determined for  the SOAR09  run,  we adjusted  the offsets  in
$\theta$  iteratively to  reach mutual  consistency between  all runs.
The remaining average differences in $\theta$ between any pair of runs
are  less than  $0\fdg2$, while  the  pixel scales  are consistent  to
within  0.2\% or  better.  We believe  that  the absolute  calibration
errors do not exceed 1\% in scale and $0\fdg5$ in angle, and that they
are likely smaller. The data  presented in this paper are possibly the
most accurate measurements of southern binaries done so far.

\subsection{Relative photometry of binary components}

The contrast of fringes in the PS $\beta = B/A$ (or the ratio of peaks
in the ACF $\beta/2$) is related to the magnitude difference between
binary components $\Delta m$,
\begin{equation}
\Delta m = -2.5 \log_{10} [( 1 - \sqrt{ 1 - \beta^2})/ \beta ] ~. 
\label{eq:dm}
\end{equation}

For small $\Delta m$, the slope of this relation is shallow, leading
to a larger error of relative photometry. Moreover, as the fringe
contrast is often under-estimated, the $\Delta m$ is over-estimated.
The positive bias on $\Delta m$ becomes significant for faint stars,
where the PS models fail. It is likely that the
background photons cause this effect, given that PS is related to the
intensity in a non-linear way (Eq.~\ref{eq:pow}). For faint stars,
the slope of the PS models $|p_1|$ is systematically less than normal,
indicating that something is wrong.

\begin{figure}
\epsscale{1.0}
\plotone{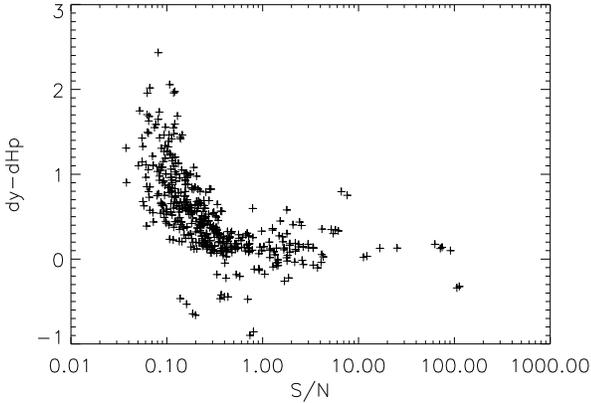}
\caption{Comparison of the magnitude difference $\Delta y$ measured by
speckle at Blanco with magnitude difference $\Delta Hp$ measured by Hipparcos
for common pairs with $\rho < 0\farcs8$, as a function of the
signal-to-noise ratio $\delta$.
\label{fig:dm-sn}}
\end{figure}

We compared our speckle photometry with that of Hipparcos \citep{HIP1997d}, 
using only the Str\"omgren $y$ data as that filter most closely matched
the Hipparcos $Hp$ filter. We found that the positive bias on $\Delta y$ is
strongly correlated with the signal-to-noise ratio $\delta$, which we
define here as the ratio of the speckle signal to the photon-noise
bias at 1/2 of the cutoff frequency,
\begin{equation}
\delta = T_0(0.5) \; 10^{p_0 + 0.5 p_1} / P_{noise} ~.
\label{eq:sn}
\end{equation}
Figure~\ref{fig:dm-sn} shows such correlation for the Blanco run.
Observations with $\delta < 0.25$ are marked by colons in the data
table. These biased estimates are still useful as upper limits,
especially for close pairs where no other photometry is available.

Another reason for $\Delta m$ over-estimation is the loss of
correlation between speckle patterns of wide binary components
(anisoplanatism). We implemented an alternative scheme for estimating
$\Delta m$ directly from average re-centered images, provided that the
binary is resolved ($\rho$ larger than the half-width of long-exposure
image). The relative position of the components is already known from
the PS fitting, so we have to determine only the true quadrant and
$\Delta m$. The quadrant is selected by comparing two point-spread
functions (PSFs) obtained by de-convolving the average image from the
binary. For the wrongly selected quadrant, the PSF has a negative lobe
opposite to the companion, so we select the quadrant with the least
negative PSF. The de-convolved PSFs are then computed for a grid of
$\Delta m$ values from 0 to the $\Delta m$ estimated from speckle. The final
$\Delta m$ value is the one which gives the most symmetrical PSF, when
the secondary component disappears. This procedure is applied
automatically to all data, but in some cases (wrong quadrant choice
for $\Delta m \sim 0$ or large $\Delta m$) it fails. In the
following, we call this method {\it resolved photometry} and mark such
$\Delta m$ estimates to distinguish them from the standard speckle
photometry of closer pairs.
 
\begin{figure}
\epsscale{0.8}
\plotone{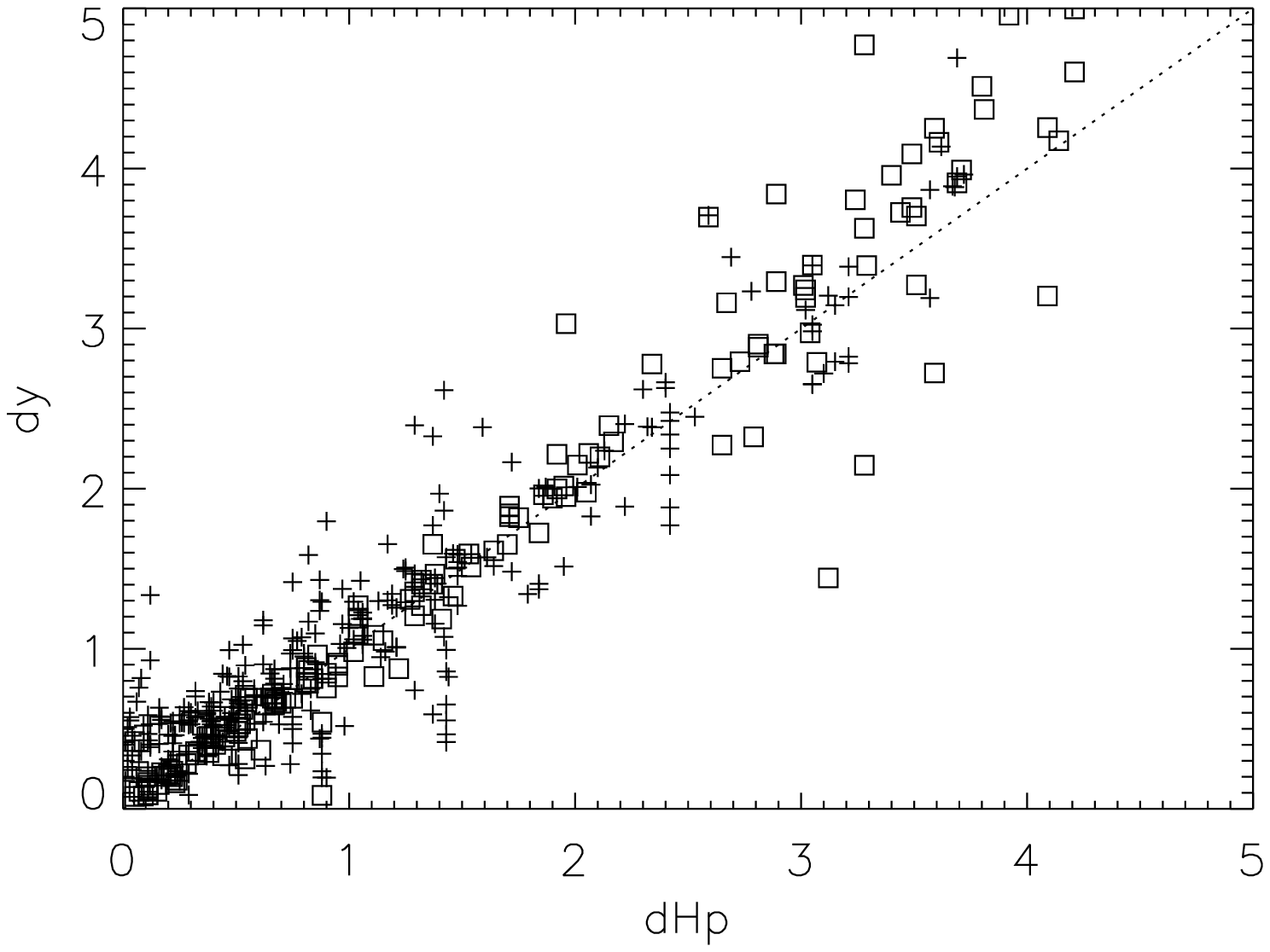}
\plotone{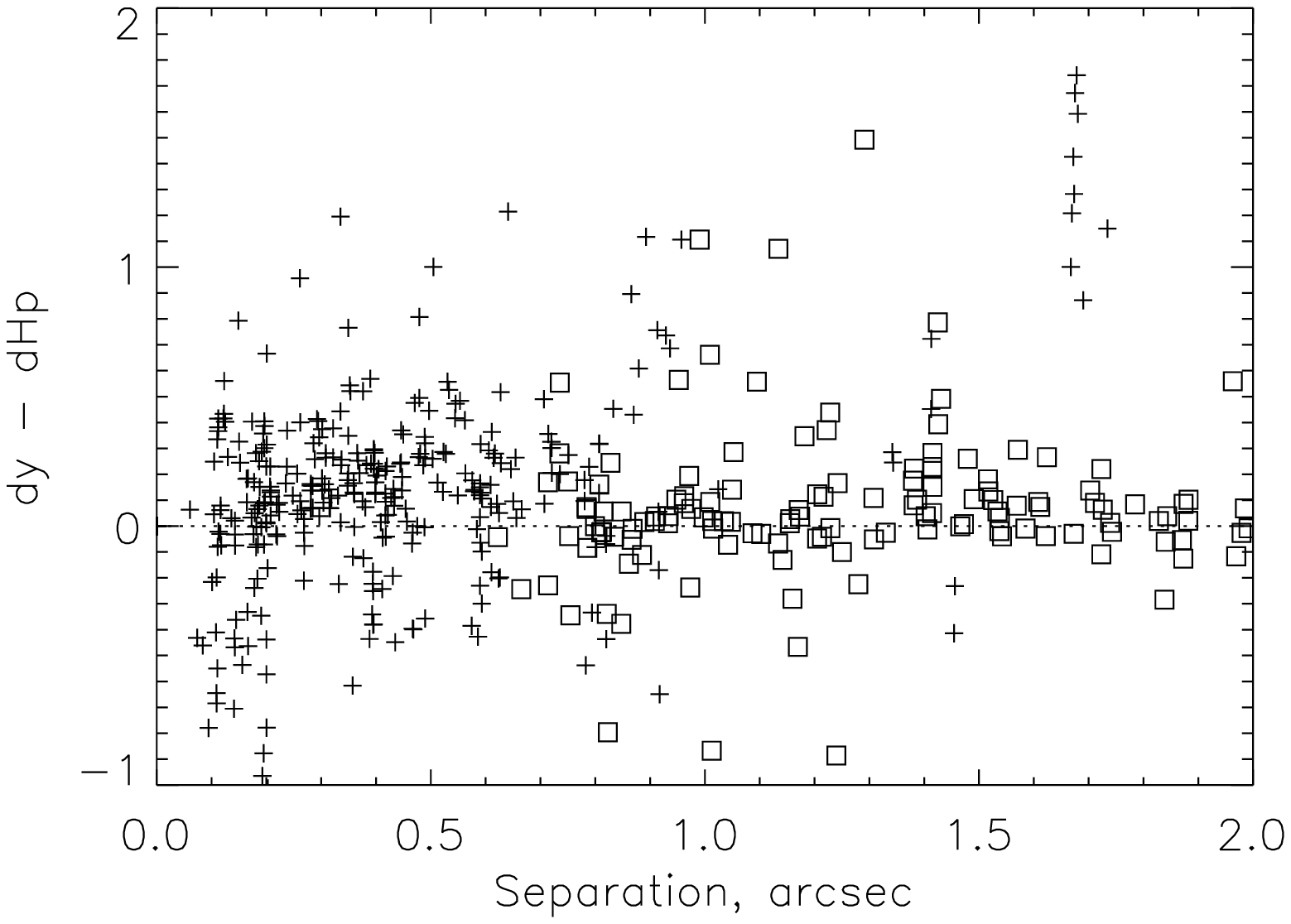}
\caption{Comparison of the magnitude difference $\Delta y$ measured by
speckle (crosses, $\delta > 0.25$) and by resolved photometry (squares)
with magnitude difference $\Delta Hp$ measured by Hipparcos for the
whole data set. Top: common pairs with $\rho < 1''$, bottom:
$\Delta y -\Delta Hp$ as a function of separation $\rho$.
\label{fig:dm-rho}}
\end{figure}

Figure~\ref{fig:dm-rho} compares $\Delta y$ measured by speckle (only
data with $\delta > 0.25$) and by resolved photometry with the
Hipparcos photometry, for the whole data set. These plots help to
evaluate the accuracy of our photometry. Despite some remaining
deviant points, the overall agreement is evident. Wide pairs without
resolved photometry suffer from the positive $\Delta y$ bias caused by
anisoplanatism. As this bias is variable, depending on high-altitude
turbulence, we do not attempt to quantify it. 

Quantitative evaluation of the bias and precision of our photometry is
given in Table~\ref{tab:ptm}. We compare speckle photometry of close
($\rho < 1''$) pairs with good S/N ($\delta > 0.25$) and resolved
photometry of wide pairs with magnitude differences $\Delta Hp$
measured by Hipparcos (first two lines) and with magnitude differences
$\Delta V$ derived from the Tycho data by \citet{FM2000} (last two
lines). Each line lists the number $N$ of pairs in common, median and
average difference between $\Delta m$'s, and the rms dispersion of the
difference. The speckle photometry has a small positive bias, while
the resolved photometry is essentially unbiased. The speckle
photometry bias is larger in comparison with the Tycho $\Delta V$ than
with $\Delta Hp$ because only pairs wider than 0\farcs3 are considered
by \citet{FM2000}, hence larger contribution of anisoplanatism. When
we compare our $\Delta y$ with $\Delta Hp$ only for pairs with $0\farcs
3 < \rho < 1''$, a similar positive bias of $0.2^m$ appears, but the
rms scatter becomes smaller, around $0.20^m$. The Hipparcos photometry
of close pairs with $\rho < 0\farcs2$ (below the resolution limit of
the Hipparcos telescope) is suspect, as can be seen in the lower plot
of Fig.~\ref{fig:dm-rho}. We estimate that intrinsic random errors of
our photometry (both speckle and resolved) are around $0.2^m$
r.m.s., but in a few cases the disagreement between our and published 
photometry is much larger.

\begin{table}
\center
\caption{Comparisons of relative photometry}
\label{tab:ptm}
\medskip
\begin{tabular}{l c c c c} 
\tableline\tableline
Data & $N$ & Med.  & Aver. & r.m.s. \\
     &     & (mag) & (mag) & (mag)  \\
\tableline
Sp.  $\Delta y - \Delta Hp$          & 303 & 0.13 & 0.11 & 0.34 \\
Res. $\Delta y - \Delta Hp$          & 162 & 0.04 & 0.08 & 0.42 \\ 
Sp.  $\Delta y - \Delta V_{\rm Tyc}$ & 206 & 0.23 & 0.28 & 0.30 \\
Res. $\Delta y - \Delta V_{\rm Tyc}$ & 132 & 0.02 & 0.05 & 0.19 \\
\tableline
\end{tabular}
\end{table}

\subsection{Detection limits}

\begin{figure}
\epsscale{1.0}
\plotone{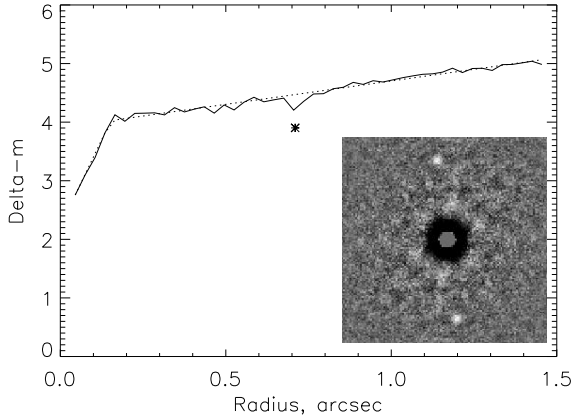}
\caption{Fluctuations in the filtered ACF of a faint binary star are
calculated, translated to the $3 \sigma$ detection limit $\Delta
m_3$ and plotted as a function of radius. The asterisk marks $(\rho,
\Delta m)$ of the actual binary companion. Approximation of the
curve by two linear segments is plotted as a dotted line. The insert
shows a fragment of the ACF, with the central peak masked.
\label{fig:detlim}}
\end{figure}

Binary companions are detected as symmetric spikes in the filtered
ACF. Fluctuations in the ACF caused by photon noise, residual
speckle statistics, etc. prevent detection of faint companions.
The rms fluctuations in the annuli of 2-pixel width around the central
peak are calculated for each ACF; this $\sigma(\rho)$ curve is
translated to the detection limit $\Delta m_3 (\rho)$ by assuming that
all companions above $3 \sigma$ are detectable. Of course, the
annulus containing the actual companion will have enhanced
fluctuations and lower $\Delta m_3$. However, as shown in
Fig.~\ref{fig:detlim}, a simple linear model can be fitted to the
$\Delta m_3 (\rho)$ curve by excluding the companion zone. Typically,
the slope of the curve changes abruptly at some distance $\rho^* \sim 0.15''$,
so we approximate it by two linear segments intersecting at
$\rho^*$. Such 2-segment linear models are fitted to all data. 
 
\begin{figure}
\epsscale{1.0}
\plotone{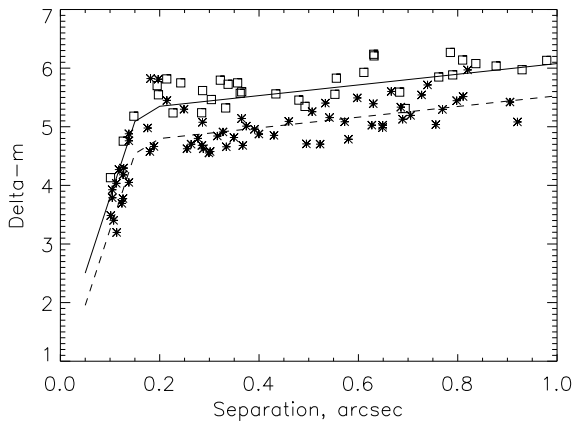}
\caption{Verification of the companion detection limit (example).
Solid and dashed lines indicate the $\Delta m_3$ and $\Delta m_5$
curves, respectively. Positive detections are plotted as asterisks,
failed detections as squares.
\label{fig:detcheck}}
\end{figure}

The detection threshold was checked by simulating fake companions. A
real ACF of a single star (or a binary de-convolved from a faint
companion) was used as a model of the speckle PSF, then companions
were generated with separations from 0\farcs1 to $1''$ and $\Delta m$
in the $(-1.0, +0.5)$ interval around $\Delta m_3$. About 10
representative cases were tested in this way, with 100 trial
companions each (Fig.~\ref{fig:detcheck}). Our general conclusion is
that the $5\sigma$ line $\Delta m_5 = \Delta m_3 - 0.55$ corresponds
to certain detection, companions in the region between $\Delta m_5$
and $ \Delta m_3$ are detected fairly frequently, and companions
with $ \Delta m > \Delta m_3$ remain undetected, with few exceptions.

\begin{figure}
\epsscale{1.0}
\plotone{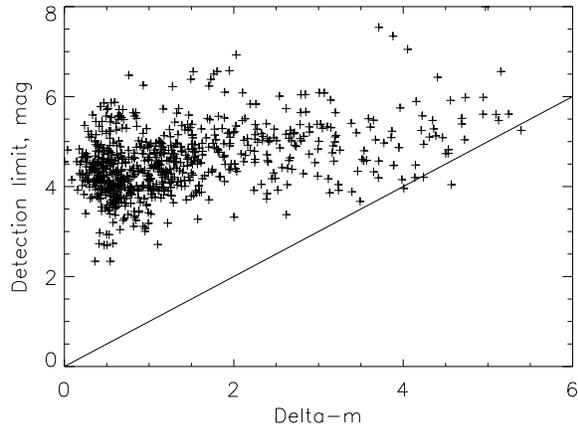}
\caption{Comparison of the measured $\Delta m$ (horizontal axis) with
the detection limits (vertical axis). Only those 777 pairs with $\rho >
0\farcs15$ and valid speckle photometry are plotted.
\label{fig:dmlim}}
\end{figure}

Figure~\ref{fig:dmlim} compares the detection limits estimated by the
above procedure with the actually measured $\Delta m$. Only data with
good signal-to-noise $\delta > 0.25$ and $\rho > 0\farcs15$ are
selected. Positive bias in $\Delta m$ inherent to speckle photometry
is also relevant to the detection limits which are over-estimated by
the same amount. For wide companions with $\rho > 1''$ anisoplanatism becomes
important, making our formal detection limits optimistic.
 The same is true for the noisy data with $\delta <
0.25$. We list the $\Delta m_5$ detection limits for unresolved
targets at separations of 0\farcs15 and $1''$ and mark cases with
$\delta < 0.25$ by colons. The actual detection limits for companions
closer than $0.1''$ cannot be established by the above simple
analysis, as they depend on a number of artifacts discussed in the
next sub-section. Median detection limits $\Delta m_5$ for the whole
data set are $4.22^m$ and $5.33^m$ at 0\farcs15 and $1''$,
respectively. For the best 25\% of data, these limits exceed $4.67^m$
and $6.08^m$.

\subsection{Artifacts and false companions}

In all data sets, most ACFs are round, but some ACFs show symmetric
enhancements near the first diffraction ring which can be mistaken
for a binary companion with $\Delta m \sim 3$. In the case of binary
stars, these false details appear around the secondary peaks as well,
distinguishing them from true triple systems where the secondary peaks
are doubled, not tripled.

These false peaks have some common features. First, their separation
from the center, typically from 45\,mas to 75\,mas, is larger in the
H$\alpha$ filter than in the y filter, while the intensity of the
peaks is also larger in H$\alpha$. Secondly, the peaks are almost
always oriented vertically, along the AD direction. Yet, they are not
caused by the AD because the separation does not depend on the
spectral bandwidth and the zenith distance (the peaks are seen even
near the zenith). Third, the separation, orientation, and intensity
of the peaks is variable. In two data cubes of the same star, one
may have the false peaks while the other does not. When we split the
data cube into 10 segments and calculate the PS for each segment, the
variability of the peaks on a time scale of seconds becomes even more
apparent. However, the peaks often appear persistently in the ACFs of
different objects observed one after another in the same part of the
sky.

The persistent nature of these peaks means that they are not caused by
random fluctuations of speckles and do not disappear when more data
are averaged. Having at least two data cubes for each object and
examining data on other objects observed before and after the star
usually helps to identify and reject false companions, despite their
striking resemblance to real binaries in some cases.

\begin{figure}
\epsscale{1.0}
\plotone{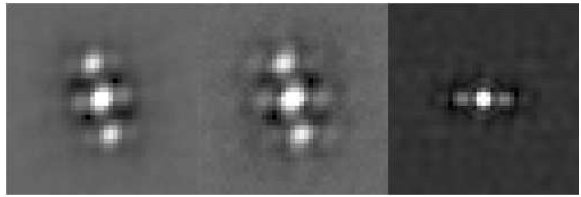}
\caption{Illustration of false peaks in the ACFs. Central $0\farcs6
\times 0\farcs6$ portions of the filtered ACFs of the binary star
BU~368AB ($\rho = 0\farcs12$) observed in August 2008 in the filters $y$
(left) and H$\alpha$ (center) are shown. The right panel shows the
simulated ACF of a single star in H$\alpha$ with optical aberrations
of $\frac{D}{2}$ period and amplitude 0.8\,$\mu$m, under $0\farcs8$ seeing, and
on the same scale. All images have the same square-root intensity
stretch from minimum (black) to 0.5 of the ACF maximum (white).
\label{fig:ears}}
\end{figure}

The properties of the false peaks indicate that they are likely caused
by variable optical aberrations with characteristic size of 2\,m, or
\onehalf ~of the telescope diameter. We simulated speckle data by adding a
sinusoidal wave-front aberration with $\frac{D}{2}$ period to the atmospheric
distortions. Some characteristics of the false peaks
(Fig.~\ref{fig:ears}, right) could be reproduced. The intensity of
the false peaks is larger in H$\alpha$ than in y, and it decreases
with degrading seeing. Orientation of the peaks in the vertical
direction suggests that optical aberrations such as astigmatism could
play some role, but our simulations show that the STF can be affected
only by a fairly large amount of defocus and astigmatism causing
visible elongation of the seeing-limited PSF. Even then the
astigmatism produces an elongation of speckles, rather than their
tripling. Air stratification in the dome can possibly cause this
optical effect, but its exact nature remains mysterious. Such false
peaks could explain previous detections of speckle companions which
turned out to be bogus. 
See the discussion of false speckle companions by \citet{McA1993}. 

Sometimes speckle peaks in the ACF are also elongated at a large angle
with respect to the AD. This blur could be caused by telescope
aberrations or tracking errors. Although tracking errors are usually
slow (typical frequency 1\,Hz), their amplitude can be large enough to
degrade the resolution in a 20-ms exposure.
Alternatively, speckles can be elongated by fast
turbulence. Instrumental elongation of speckles is indistinguishable
from the effect of a close binary companion at the limit of resolution
($\sim 30$\,mas), so only the examination of data on stars observed
before or after can help to distinguish an authentic close binary from an
artifact.

Clearly, detection and measurement of close binary companions are
complicated by the artifacts and involve an element of human judgment
and error. The detection limits cannot be formalized, as was done for
wider companions. We cannot exclude the possibility that some of the measurements
presented below are affected by the artifacts, despite all efforts to
understand and eliminate them.

\section{Results}

\subsection{Data tables}


Table~\ref{tab:double} lists 1898 measurements of 1189 resolved known and new binary
stars and sub-systems. Its columns contain (1) the WDS \citep{WDS} 
designation, (2) the ``discoverer designation" as adopted in the WDS, (3) an
alternative name, mostly from the Hipparcos catalog, (4) Besselian epoch of 
observation, (5) filter, (6) number of individual data cubes, (7,8) position
angle $\theta$ in degrees and internal measurement error in tangential direction 
$\rho \sigma_{\theta}$ in mas, (9,10) separation $\rho$ and its
internal error 
$\sigma_{\rho}$ in mas, and (11) magnitude difference $\Delta m$. An 
asterisk follows the value if $\Delta m$ and the true quadrant are 
determined from the resolved photometry; a colon indicates that the data are
noisy and $\Delta m$ is likely over-estimated. We decided not to mark with 
colons the $\Delta m$ values of wide pairs over-estimated only due to 
anisoplanatism (when no resolved photometry is available), to avoid 
confusion with the low S/N cases. Note that in the cases of multiple stars, 
the positions and photometry refer to the pairings between individual stars,
not with photo-centers of sub-systems.

For stars with known orbital elements, columns (12--14) of 
Table~\ref{tab:double} list the residuals to the ephemeris position and the 
reference to the orbit from the {\it 6$^{th}$ Orbit Catalog} \citep{VB6}. In
those cases where multiple orbits for the same system are present in the 
catalog, the orbit with the smallest residuals is selected. An asterisk in 
the final column indicates that a note concerning this system may be found 
in Table~\ref{tab:notes}.

Table~\ref{tab:single} contains the data on 285 unresolved stars, some of 
which are listed as binaries in the WDS or resolved here in other runs. Columns (1) through (6) are the 
same as in Table~\ref{tab:double} (although Column (2) also includes Bayer 
designations, HD numbers, or other names for objects without discoverer 
designations). Columns (7,8) give the $5 \sigma$ detection limits 
$\Delta m_5$ at $0\farcs15$ and $1''$ separations determined by the 
procedure described above. When two or more data cubes are processed, the 
best detection limits are listed. Noisy data with $\delta < 0.25$ are marked
by colons to indicates that the actual detection limits are smaller. As in 
Table~\ref{tab:double}, the final column indicates a note to the system.

New discoveries are repeated  in Table~\ref{tab:new} in the same 
format as measurements in Table~\ref{tab:double} -- a total of 48 pairs. 
Figure~\ref{fig:triple} shows ACFs of 20 newly resolved triple systems.

\subsection{Comments on individual objects}

Notes to some objects in Tables~\ref{tab:double}, \ref{tab:single}, and 
\ref{tab:new} are given in Table~\ref{tab:notes}. These notes include 
miscellaneous information such as additional components, discovery history, 
etc. The WDS \citep{WDS} and the Multiple-Star Catalog \citep{MSC} were 
extensively consulted, among other sources. Each system is identified by 
its WDS designation and an alternate name. Cases where deviations from the orbits 
are quite large are also indicated. The definition of unacceptably large 
residuals is subjective; we consider as such orbits which deviate from our 
measurements by more that $20^\circ$ in $\theta$ or by more than 50\% in $\rho$. 
There are 131 such cases out of 544 systems with orbits. The 24\% fraction of 
{\it bad orbits} demonstrates that more measurements of southern binaries are needed.

In this sub-section we give more lengthy comments on a few selected cases.

{\bf 02053-2425 = HIP~9774 = I~454:} The brightest companion of I~454
(also known as ADS~1652) is a double-lined spectroscopic binary with
period 2.6\,yr and eccentric orbit, $e= 0.78$ (Tokovinin,
unpublished). The estimated semi-major axis of this pair is 47\,mas.
The system passed through periastron in May-June 2008 and was
marginally resolved in July 2008 at Blanco. In October 2008 the
separation was closer, below the diffraction limit of the 4-m
telescope. Nevetheless, we were able to fit consistently a
triple-star model to 7 power spectra recorded in October. As our
estimated $\Delta m \sim 1^m$ is larger than the spectroscopically
estimated $\Delta m \sim 0.3^m$, it is possible that the actual
separations were even smaller than those listed in
Table~\ref{tab:new}. Component C = HIP~9769 of this
multiple system was also observed here and found to be single.

{\bf 02225-2349 = HIP~11072 = $\kappa$~For = TOK~40:} 
The companion was first resolved in 2007 \citep{TC08}. New
measurements are roughly compatible with the 26.5-yr astrometric orbit
of \citet{GK02} if we adjust the semi-major axis to $0\farcs65$ and the
node position angle to $120^\circ$. New photometry ($\Delta y = 5.0$,
$\Delta {\rm H} \alpha = 4.3$) shows that the companion is not as
faint as measured initially, and that it is redder than the primary
star. It will be very useful to measure the relative brightness of the
companions in the near-IR with adaptive optics.

{\bf 05086-1810 = HIP~23932 = WSI~72:} Speculated to be a close binary by 
\citet{Hen2002}, it was first resolved
in 2006 with the USNO speckle camera on the Blanco 4m by \citet{GDwarf} at
about the same position angle and twice the separation measured here.

{\bf 05354-0555 = HIP~26241 = CHR~250Aa,Ab:} New measurements confirm
slow rectilinear motion of this pair due either to a
long-period orbit or to Ab being an unrelated background star, in
agreement with the conclusions of \citet{Msn2009}. Our photometry
indicates that CHR~250Ab is bright, V$ \sim 6^m$.

\begin{figure}[t]
\epsscale{1.0}
\plotone{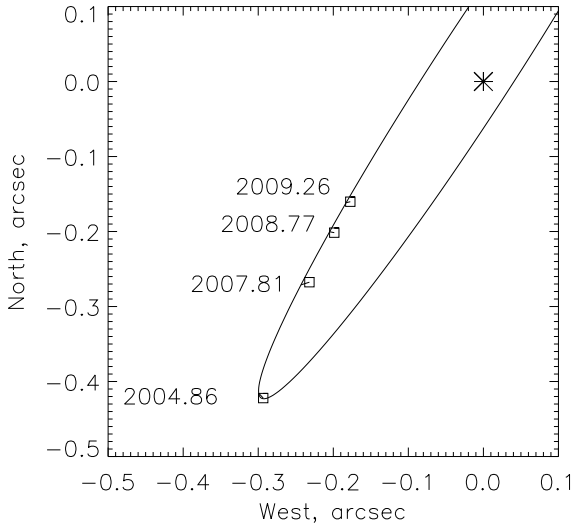}
\caption{Observed motion of the E component (squares) relative to the
 C component (big star) in the CE sub-system of the visual
 quadruple star GJ~225.2 The separation decreases from 0\farcs514 in
 2004.86 to 0\farcs239 in 2009.26. The line shows a possible orbit
 with 23.7-year period.
 \label{fig:Tok9}}
\end{figure}

{\bf 06003-3102 = HIP~28442C = GJ~225.2C = TOK~9CE:} 
The E companion in this nearby quadruple system was discovered
with adaptive optics in 2004 \citep{Gl2252}. In that paper, a
tentative astrometric orbit with 23.7-yr period was proposed, and an
unusually ``blue'' $J-K$ color of E was noted. The component E was
marginally resolved in the visible during the first speckle run at SOAR
\citep{TC08}. In the present data, the companion is seen reliably
above the detection limit, and its magnitude difference is measured
consistently as $\Delta y = 4.5^m$. All 4 position measurements
available to date show retrograde orbital motion (Fig.~\ref{fig:Tok9})
which does not follow the preliminary astrometric orbit suggested by
\citet{Gl2252}, but is still compatible with a 24-year orbital
period. Within a few years, the orbit can be established more firmly
and we will be able to address the properties of this apparently
peculiar companion. New observations of the other sub-system AB
confirm its orbital elements.

{\bf 06410+0954 = HIP~31978 = 15 Mon = CHR~168Aa,Ab:}
The most recent orbit of \cite{Gie1997} is clearly in error. This system
has been the subject of regular observation by both speckle interferometry
and HST-FGS. A new orbit is currently in preparation.

\begin{figure*}
\epsscale{2.0}
\plotone{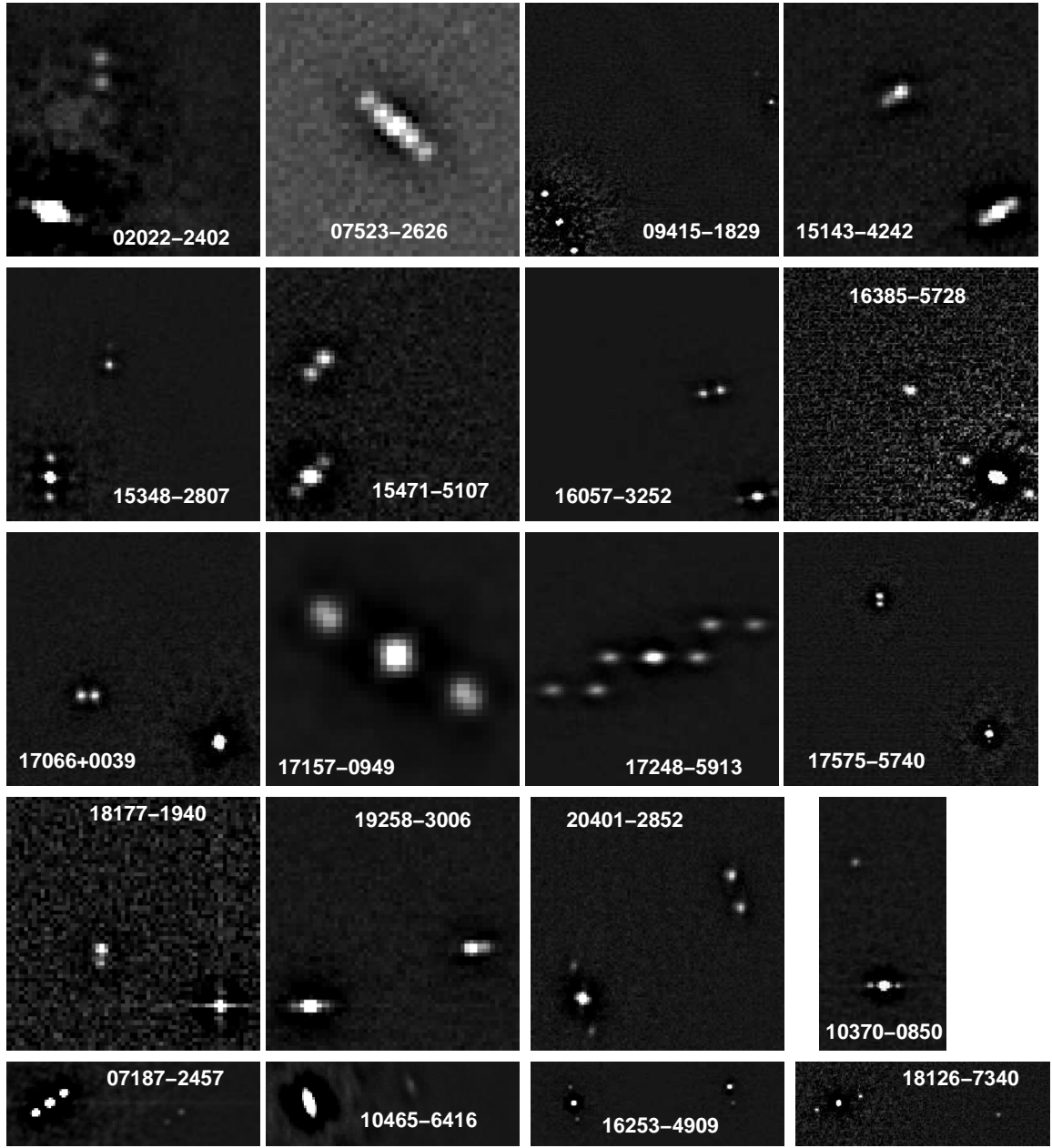}
\caption{ACFs of 20 triple stars with newly resolved components.
Each panel shows a fragment of the filtered ACF including the central
peak and some or all companion peaks, in arbitrary intensity stretch and 
with detector lines horizontal (not necessarily North up).
 \label{fig:triple}}
\end{figure*}

{\bf 07523-2626 = HIP~3840 = V402 Pup = WSI 54:}
This $9.23^m$ star of spectral type O6e belongs to the open cluster
NGC~2467. \citet{Msn2009} have resolved it into a close pair WSI~54
and measured in 2006.194 the position angle $231.8^\circ$ and the
separation $0\farcs091$. Our observations clearly show three stars in
a tight linear configuarion (Fig.~\ref{fig:triple}). The outer
companion matches the WSI~54 pair best, the inner companion has a
separation 2 times smaller and a similar flux. This detection is based
on 3 independent data cubes, the companions are not aligned with the
AD, so we are confident that this is not an artifact. Further
observations will reveal whether this is a dynamically unstable system
(trapezium), an unsual multiple with orbits in resonance, or a chance
projection of a binary and a single star, more probable in a cluster
than in the field.

{\bf 17248-5913 = HIP~85216 = I~385 + WSI 85:} 
Quite unexpectedly this star, previously considered as a binary,
turned out to be a spectacular triple with components of comparable
magnitude and separation (trapezium-type); see Fig.~\ref{fig:triple}.
During the 0.72-yr time between the Blanco and SOAR09 runs, the relative
position of both companions changed only slightly; the magnitude
differences remained stable as well. Comparing our measurements with
published data, we identify the wider companion at $122^\circ$,
$0\farcs39$ with the previously known component B and designate the new
companion at $270^\circ$, $0\farcs26$ as D. The distant companion C at
$210^\circ$, $17''$ (also seen in the 2MASS images) is $5^m$ fainter
than A and has moved only slightly since its discovery in
1901. Therefore, C likely belongs to this system.

The AB pair has revolved by $64^\circ$ over the 110 years since its 
discovery by \citet{Innes05}, suggesting an orbital period of $\sim 600$~yr; 
this corresponds also to the dynamically estimated period at a distance of 
211\,pc measured by Hipparcos. The projected separation of AD (55\,AU) means
that its orbital period should be of the order of 300\,yr. 

The component D was noted  in the only previous speckle observation by
\citet{Hart93},  but  it  was  not  accepted as  real  at  that  time.
Re-measurement  of the  correlation  peak corresponding  to AD  yields
$0\farcs199, \; 270\fdg18$ on 1990.3496.  It seems that the pair AD is
slowly opening up. The AD was not resolved visually at a 1-m telescope
by \citet{Holden77}.  The Hipparcos measured the  relative position of
AB at  $118^\circ$, $0\farcs452$ in disagreement with  all other data,
as  though the  light center  of AD  was measured  instead of  A.  The
mysterious new companion D  deserves futher observations.

{\bf 17535-0355 = HD~162905 = V2610 Oph = TOK 54:} According to \citet{Pri2009},
this is a close quadruple system composed of two double-lined binaries
with periods 8.47 and 0.425 days (the latter is also eclipsing)
orbiting each other. All components are dwarfs of
spectral types F to G. The outer system is resolved here for the
first time. The authors estimate $\Delta m = 0.27^m$; our measurement
$\Delta y = 1^m$ is biased by the low S/N. 
Some other eclipsing binaries discovered to be
spectroscopic multiples by the same team were also observed at SOAR in
April 2009 and found unresolved. However, one of those, HDS~238, has a
visual companion at 3\farcs2 listed in the WDS, too wide to
be measured here.

{\bf 18455+0530 = HIP~92027 = STF~2375AB, FIN~332Aa,Ab \& FIN~332Ba,Bb:}
New orbits, based on new reductions of historical interferometric
measures and new measures, are in process for this complex
multiple system \citep{Tweedle}.

{\bf 20401-2852 = HD~196718 = SEE~423:} This triple system ($V=8.70$, F5V) 
was first resolved in 1897 at $(0\farcs92, 20^\circ$). Three observations of
this pair are listed in Aitken's (1932) catalog under \#14115, showing slow 
direct motion. \citet{Holden77b} measured in 1976.8 $(0\farcs76, 32^\circ)$ 
and $\Delta m =0.5$. However, a larger separation of 
$(1\farcs112, 33^\circ)$ was measured by the Tycho experiment on 1991.68 
\citep{HIP1997d}. One year later, in 1992.4552, \citet{Hart96} found the 
pair at a very different position, $(0\farcs34, 271^\circ)$. We see now that
this system is a visual triple (Fig.~\ref{fig:triple}), with the wide pair 
AB at $(1\farcs15, 38^\circ)$ matching the Tycho result and corresponding to
SEE~423. The closer pair BC is fainter than A by $0.64^m$ (Tycho) and was 
apparently measured for the first time in 1992 by \citet{Hart96}.
The triple nature of SEE~423 clarifies some, but not all contradictions in 
the existing data. The wider (and the brightest) component was not seen by 
speckle in 1992 as it was outside the field-of-view. However, why did Holden
and other visual observers not resolved the sub-system BC with nearly equal 
components? Also, why is the measured separation of the wide pair so 
discordant, ranging from 0\farcs76 (Holden) to 1\farcs15 (this work)?


\acknowledgments

We are grateful to operators of the Blanco telescope A.~Alvarez,
H.~Tirado, C.~Aguilera and to the operators of SOAR D.~Maturana, 
S.~Pizarro, P.~Ugarte, A.~Past\'en for their dedicated and efficient 
work enabling observations of so many stars per night. The 
development of HRCam software by R.~Cantarutti was essential for 
this project. Comments by anonymous Referee helped to improve the
presentation. 

The USNO speckle interferometry program has been supported by the National Aeronautics and 
Space Administration under Grant No. NNH06AD70I, issued through the Terrestrial Planet 
Finder Foundation Science program. This research has made use of the SIMBAD database,
operated at CDS, Strasbourg, France. Thanks are also extended to Ken Johnston and the 
U.\ S.\ Naval Observatory for their continued support of the Double Star Program.

{\it Facilities:} \facility{Blanco}, \facility{SOAR}.







\clearpage

\begin{deluxetable}{l l l  ccc  rc cc l r r l c}                                                                                                                                
\tabletypesize{\scriptsize}                                                                                                                                                     
\rotate                                                                                                                                                                         
\tablecaption{Measurements of known and new binary stars (fragment)                                                                                                                                
\label{tab:double} }                                                                                                                                                            
\tablewidth{0pt}                                                                                                                                                                
\tablehead{                                                                                                                                                                     
\colhead{WDS} & \colhead{Discoverer} & \colhead{Other} & \colhead{Epoch} & \colhead{Filt} & \colhead{N} & \colhead{$\theta$} & \colhead{$\rho \sigma_{\theta}$} &               
\colhead{$\rho$} & \colhead{$\sigma \rho$} & \colhead{$\Delta m$} & \colhead{[O$-$C]$_{\theta}$} & \colhead{[O$-$C]$_{\rho}$} & \colhead{Reference} & \colhead{Note} \\         
\colhead{(2000)} & \colhead{Designation} & \colhead{name} & +2000 & & & \colhead{(deg)} & (mas) & ($''$) & (mas) & (mag) & \colhead{(deg)} & \colhead{($''$)}                   
& \colhead{code$^*$} & }                                                                                                                                                        
\startdata                                                                                                                                                                      
00006$-$5306 & HJ  5437       & HIP     50    & 8.7697 & y         & 2 & 332.7 &  0.2 &  1.4904 &  0.7 & \phm{$-$}3.4~~* &                 &                   &          &   \\
00008$+$1659 & BAG   18       & HIP     68    & 8.5378 & y         & 1 &   6.8 &  6.0 &  0.6112 &  6.1 & \phm{$-$}4.6    &                 &                   &          &   \\
00008$-$3244 & I   1478       & HD  224811    & 8.5459 & y         & 2 & 328.6 &  0.3 &  0.3766 &  0.3 & \phm{$-$}1.1~~: &                 &                   &          &   \\
00028$+$0208 & BU   281 AB    & HIP    223    & 8.5379 & y         & 5 & 161.9 &  0.2 &  1.5689 &  1.0 & \phm{$-$}2.0~~* &                 &                   &          & * \\
             &                &               & 8.7672 & y         & 2 & 161.7 &  0.5 &  1.5709 &  0.2 & \phm{$-$}2.2~~* &                 &                   &          &   \\
00039$-$5750 & I    700       & HIP    306    & 8.5459 & y         & 2 & 144.1 &  0.2 &  0.2965 &  0.2 & \phm{$-$}0.8~~: &                 &                   &          &   \\
00059$+$1805 & STF 3060 AB    & HIP    495    & 8.5377 & y         & 2 & 133.6 &  1.5 &  3.4185 &  1.5 & \phm{$-$}0.3~~* &                 &                   &          & * \\
00059$-$3020 & RST 5180 AB    & HD     117    & 8.5459 & y         & 2 & 340.3 &  0.3 &  0.3274 &  0.3 & \phm{$-$}1.2~~: &                 &                   &          & * \\
00090$-$5400 & HDO  181       & HIP    730    & 8.5379 & y         & 2 &  35.4 &  0.3 &  0.3236 &  0.1 & \phm{$-$}1.8    &   $-$3.4\phm{X} &   $-$0.013\phm{X} & Alz2000b &   \\
             &                &               & 8.5379 & H$\alpha$ & 2 &  35.4 &  0.2 &  0.3228 &  0.2 & \phm{$-$}2.0    &   $-$3.4\phm{X} &   $-$0.013\phm{X} & Alz2000b &   \\
             &                &               & 8.5431 & y         & 2 &  35.5 &  0.4 &  0.3228 &  0.4 & \phm{$-$}1.7~~: &   $-$3.3\phm{X} &   $-$0.013\phm{X} & Alz2000b &   \\
             &                &               & 8.5431 & H$\alpha$ & 2 &  35.3 &  0.7 &  0.3223 &  0.8 & \phm{$-$}2.2~~: &   $-$3.5\phm{X} &   $-$0.014\phm{X} & Alz2000b &   \\
             &                &               & 8.5486 & y         & 3 &  35.3 &  0.1 &  0.3237 &  0.3 & \phm{$-$}1.6    &   $-$3.5\phm{X} &   $-$0.013\phm{X} & Alz2000b &   \\
             &                &               & 8.6059 & y         & 2 &  35.2 &  0.1 &  0.3241 &  0.2 & \phm{$-$}1.6    &      9.8\phm{X} &   $-$0.060\phm{X} & Sey2001  &   \\
00098$-$3347 & SEE    3       & HIP    794    & 8.5459 & y         & 2 & 116.1 &  0.3 &  0.7827 &  0.3 & \phm{$-$}1.5~~: &     38.5\phm{X} &      0.100\phm{X} & Csa1983a & * \\
             &                &               & 8.5459 & H$\alpha$ & 2 & 116.1 &  0.7 &  0.7819 &  0.7 & \phm{$-$}1.5~~: &     38.5\phm{X} &      0.099\phm{X} & Csa1983a &   \\
             &                &               & 8.5486 & y         & 3 & 116.1 &  0.5 &  0.7830 &  0.7 & \phm{$-$}1.7~~: &     38.5\phm{X} &      0.100\phm{X} & Csa1983a &   \\
00115$-$5545 & HDS   25       & HIP    927    & 8.5459 & y         & 2 &  77.5 &  0.4 &  0.1775 &  0.2 & \phm{$-$}1.0    &                 &                   &          &   \\
00121$-$5832 & RST 4739       & HIP    975    & 8.5431 & y         & 2 & 134.8 &  0.6 &  0.3191 &  0.6 & \phm{$-$}1.4~~: &                 &                   &          & * \\
             &                &               & 8.7697 & y         & 2 & 133.9 &  0.1 &  0.3182 &  0.1 & \phm{$-$}0.4    &                 &                   &          &   \\
00126$-$1142 & RST 3343       & HIP   1005    & 8.5432 & y         & 2 & 253.0 &  0.4 &  0.2799 &  0.3 & \phm{$-$}1.0~~: &   $-$7.6\phm{X} &   $-$0.036\phm{X} & Hei1998  &   \\
             &                &               & 8.5487 & y         & 3 & 253.1 &  0.2 &  0.2801 &  0.2 & \phm{$-$}0.7~~: &   $-$7.5\phm{X} &   $-$0.036\phm{X} & Hei1998  &   \\
             &                &               & 8.6059 & y         & 2 & 253.1 &  0.1 &  0.2803 &  0.1 & \phm{$-$}0.4    &   $-$7.6\phm{X} &   $-$0.036\phm{X} & Hei1998  &   \\
00143$-$2732 & HDS   33       & HIP   1144    & 8.5432 & y         & 4 & 116.2 &  0.4 &  0.1737 &  0.5 & \phm{$-$}1.4~~: &                 &                   &          &   \\
\enddata
\vskip 0.1in                                                                                                                                                                    
* The complete list of references may be found at http://ad.usno.navy.mil/Webtextfiles/wdsnewref.txt .                                                                          
\end{deluxetable}                                                                                                                                                               

\begin{deluxetable}{l l l   c c c  c c c c}                                                     
\tabletypesize{\scriptsize}                                                                     
\tablecaption{Unresolved stars (fragment)                                                                  
\label{tab:single} }                                                                            
\tablewidth{0pt}                                                                                
\tablehead{                                                                                     
WDS (2000) & \colhead{Discoverer} & \colhead{Hipparcos} & \colhead{Epoch} & \colhead{Filter} &  
\colhead{N} & \multicolumn{2}{c}{5$\sigma$ Detection Limit} & $\Delta m$ & Note  \\             
 & \colhead{Designation} & \colhead{or other} & \colhead{+2000} & & &                           
\colhead{$\Delta m (0\farcs15)$}  & \colhead{$\Delta m (1'')$}    & flag &    \\             
 & \colhead{or other name}  & \colhead{name}  & & & & \colhead{(mag)} &  \colhead{(mag)}  &     
}                                                                                               
\startdata                                                                                      
00024$+$1100 & HD 224983      & HIP    184   & 8.5378 & y         & 2 &  4.06 &  4.63 &   &   \\
00059$+$1814 & LTT 10019      & HIP    493   & 8.5377 & y         & 2 &  4.03 &  4.49 &   &   \\
00063$-$4905 & HDO 180        & HIP    522   & 8.5379 & y         & 2 &  4.60 &  6.19 &   & * \\
00084$+$0637 & HD    377      & HIP    682   & 8.5378 & y         & 2 &  4.60 &  5.46 &   &   \\
00113$-$1528 & 6 Cet          & HIP    910   & 8.5379 & y         & 2 &  4.96 &  6.19 &   &   \\
00116$+$1020 & HD    727      & HIP    943   & 8.5378 & y         & 2 &  3.99 &  4.55 &   &   \\
00117$-$3508 & the Scl        & HIP    950   & 8.5379 & y         & 2 &  4.76 &  6.27 &   &   \\
00125$+$1434 & LN Peg         & HIP    999   & 8.5378 & y         & 2 &  3.66 &  4.19 &   &   \\
00174$+$0853 & STF  22 C      & HIP   1392   & 8.7672 & y         & 2 &  4.06 &  5.84 &   & * \\
00201$-$6452 & zet Tuc        & HIP   1599   & 8.5379 & y         & 2 &  4.60 &  6.01 &   &   \\
00258$-$7715 & bet Hyi        & HIP   2021   & 8.5379 & y         & 2 &  4.41 &  6.03 &   &   \\
             &                &              & 8.7726 & H$\alpha$ & 6 &  4.95 &  6.15 &   &   \\
00291$-$0742 & MLR   2        & HIP   2275   & 8.5487 & y         & 3 &  3.71 &  4.20 & : &   \\
00327$-$6302 & B     8 A      & HIP   2578   & 8.7697 & y         & 2 &  4.64 &  6.59 &   & * \\
00327$-$6302 & B     8 B      & HIP   2578   & 8.7697 & H$\alpha$ & 2 &  4.75 &  6.72 &   & * \\
00345$-$5222 & GJ 9017        & HIP   2711   & 8.5405 & y         & 2 &  4.73 &  6.52 &   &   \\
00366$-$4908 & LDS  21 A      & HIP   2888   & 8.5405 & y         & 2 &  4.67 &  6.25 &   &   \\
00374$-$3717 & I   705        & HIP   2944   & 8.5405 & y         & 2 &  4.68 &  6.43 &   &   \\
             &                &              & 8.5460 & y         & 2 &  4.69 &  5.53 &   &   \\
00404$-$5927 & GJ  29         & HIP   3170   & 8.5405 & y         & 3 &  4.91 &  6.14 &   &   \\
\enddata                                                                                        
\end{deluxetable}                                                                               

\begin{deluxetable}{c l l ccc rc cc l c}                                                                                                  
\tabletypesize{\scriptsize}                                                                                                               
\tablecaption{Newly resolved binary and multiple stars                                                                                    
\label{tab:new} }                                                                                                                         
\tablewidth{0pt}                                                                                                                          
\tablehead{                                                                                                                               
WDS (2000) & \colhead{Discoverer} & \colhead{Other} & \colhead{Epoch} & \colhead{Filter} & \colhead{N} & \colhead{$\theta$} &             
\colhead{$\rho \sigma_{\theta}$} & \colhead{$\rho$} & \colhead{$\sigma \rho$} & \colhead{$\Delta m$}  & \colhead{Note}  \\                
          & \colhead{Designation}  &  \colhead{name}           &  +2000          &                &             &  (deg)             &    
  (mas)                          &  ($''$)          &  (mas)                  &  (mag)               &   }                                
\startdata                                                                                                                                
01144$-$0755  & WSI  70 Aa,Ab  & HIP   5799    & 8.5405 &     y     & 2 &    111.7  &     2.3  &  0.1690  &  2.7  & \phm{$-$}4.5    & * \\
02022$-$2402  & TOK  41 Ba,Bb  & HIP   9497    & 8.7674 &     y     & 2 &      3.6  &     3.4  &  0.0886  &  3.4  & \phm{$-$}0.3    & * \\
              &                &               & 8.7674 & H$\alpha$ & 2 &      6.5  &     2.9  &  0.0914  &  7.0  & \phm{$-$}0.2    &   \\
02057$-$2423  & WSI  71 Aa,Ab  & HIP   9774    & 8.5406 &     y     & 2 &    146.7  &     3.5  &  0.0397  &  1.0  & \phm{$-$}1.7~~: & * \\
              &                &               & 8.5406 & H$\alpha$ & 2 &    143.0  &     6.5  &  0.0437  &  9.7  & \phm{$-$}2.5~~: &   \\
              &                &               & 8.7674 &     y     & 2 &    184.5  &     0.7  &  0.0266  &  0.7  & \phm{$-$}1.0    &   \\
              &                &               & 8.7727 &     y     & 3 &    206.0  &     4.9  &  0.0248  &  7.7  & \phm{$-$}1.7~~: &   \\
              &                &               & 8.7727 &     y     & 2 &    190.9  &     0.9  &  0.0217  &  1.9  & \phm{$-$}1.3~~: &   \\
05086$-$1810  & WSI  72        & HIP  23932    & 8.7677 &     y     & 2 &     47.8  &     0.9  &  0.0518  &  0.4  & \phm{$-$}0.0~~: &   \\
07187$-$2457  & TOK  42 Aa,E   & HIP  35415    & 9.2595 &     y     & 2 &     87.6  &     0.4  &  0.9480  &  1.3  & \phm{$-$}4.4    & * \\
07523$-$2626  & WSI  54 AC     & HIP  38430    & 9.2595 &     y     & 3 &    226.8  &     2.9  &  0.0450  &  0.2  & \phm{$-$}1.3~~: & * \\
09252$-$1258  & WSI  73        & HIP  46191    & 9.2597 &     y     & 2 &    274.8  &     0.6  &  0.1783  &  0.9  & \phm{$-$}1.1~~: & * \\
09415$-$1829  & TOK  43 Aa,Ab  & HIP  47537    & 9.2652 &     y     & 2 &     29.8  &     6.3  &  0.4365  &  1.4  & \phm{$-$}2.1    & * \\
              &                &               & 9.2652 &     V     & 2 &     28.2  &     4.4  &  0.4436  &  1.7  & \phm{$-$}2.4    &   \\
10370$-$0850  & TOK  44 Aa,Ab  & HIP  51966    & 9.2626 &     y     & 3 &    268.8  &     2.7  &  0.0973  &  0.3  & \phm{$-$}3.0    & * \\
10465$-$6416  & TOK  45 AC     & HIP  52701    & 9.2654 & H$\alpha$ & 4 &     11.6  &     4.5  &  0.7475  &  3.7  & \phm{$-$}3.9    & * \\
12485$-$1543  & WSI  74 Aa,Ab  & HIP  62505    & 8.5394 &     y     & 4 &    154.3  &     2.3  &  0.0461  &  0.9  & \phm{$-$}1.3    & * \\
              &                &               & 9.2599 &     y     & 3 &     99.0  &     0.3  &  0.0757  &  0.4  & \phm{$-$}1.5    &   \\
13126$-$6034  & WSI  75 Aa,Ab  & HD  114566    & 8.5448 &     V     & 2 &     75.9  &     9.7  &  0.1089  &  9.1  & \phm{$-$}2.9    & * \\
13254$-$5947  & WSI  76        & HIP  65492    & 8.5448 &     y     & 2 &    186.8  &     1.1  &  0.0949  &  1.1  & \phm{$-$}2.6~~: &   \\
              &                &               & 9.2627 &     y     & 5 &    185.3  &     2.0  &  0.1022  &  7.4  & \phm{$-$}3.2~~: &   \\
13275$+$2116  & TOK  46        & HD  117078    & 9.2601 &     y     & 4 &     23.7  &     1.1  &  0.0996  &  5.5  & \phm{$-$}1.9~~: & * \\
13513$-$2423  & WSI  77        & HIP  67620    & 9.2601 &     y     & 2 &    176.9  &     0.2  &  0.1437  &  0.2  & \phm{$-$}3.4    & * \\
              &                &               & 9.2601 & H$\alpha$ & 2 &    177.0  &     0.4  &  0.1429  &  0.2  & \phm{$-$}2.8    &   \\
13527$-$1843  & WSI  78        & HIP  67744    & 8.5395 &     y     & 2 &    100.8  &     2.0  &  0.0302  &  1.6  & \phm{$-$}0.9    & * \\
              &                &               & 9.2601 &     y     & 2 &    114.9  &     0.0  &  0.0366  &  0.1  & \phm{$-$}0.7    &   \\
              &                &               & 9.2601 & H$\alpha$ & 2 &    115.7  &     0.2  &  0.0392  &  0.5  & \phm{$-$}1.4    &   \\
14020$-$2108  & WSI  79        & HIP  68552    & 8.5394 &     y     & 4 &    149.5  &     2.0  &  0.2977  &  3.3  & \phm{$-$}2.8~~: & * \\
14581$-$4852  & WSI  80        & HIP  73241    & 8.5368 &     y     & 3 &    132.9  &     1.7  &  0.2984  &  1.6  & \phm{$-$}4.3    & * \\
              &                &               & 9.2602 &     y     & 2 &    129.0  &     0.5  &  0.3178  &  0.5  & \phm{$-$}4.4    &   \\
              &                &               & 9.2602 & H$\alpha$ & 2 &    129.2  &     0.4  &  0.3166  &  0.5  & \phm{$-$}3.7    &   \\
14589$+$0636  & WSI  81        & HIP  73314    & 9.2629 &     y     & 2 &     49.9  &     0.2  &  0.1268  &  0.5  & \phm{$-$}0.8    & * \\
14598$-$2201  & TOK  47        & HIP  73385    & 9.2628 &     y     & 2 &    171.7  &     1.5  &  0.0400  &  0.3  & \phm{$-$}1.7    & * \\
15143$-$4242  & WSI  82 Aa,Ab  & HD  134976    & 8.5477 &     y     & 2 &     30.6  &     0.0  &  0.0561  &  0.1  & \phm{$-$}1.8~~: & * \\
              &                &               & 9.2603 &     y     & 2 &     35.2  &     0.9  &  0.0572  &  3.8  & \phm{$-$}1.5    &   \\
15317$+$0053  & TOK  48        & HIP  76031    & 9.2604 &     y     & 2 &     67.9  &     2.0  &  0.0382  &  0.2  & \phm{$-$}1.2    & * \\
              &                &               & 9.2604 & H$\alpha$ & 2 &     69.6  &     2.2  &  0.0416  &  0.2  & \phm{$-$}1.0    &   \\
              &                &               & 9.2658 & H$\alpha$ & 1 &     85.4  &     0.1  &  0.0374  &  0.1  & \phm{$-$}1.0    &   \\
              &                &               & 9.2658 &     y     & 2 &     91.6  &     5.5  &  0.0393  &  2.5  & \phm{$-$}1.1    &   \\
15348$-$2807  & TOK  49 Aa,Ab  & HIP  76275    & 9.2657 &     y     & 2 &    181.0  &     0.3  &  0.1376  &  4.5  & \phm{$-$}2.3    & * \\
              &                &               & 9.2657 &     R     & 2 &    180.4  &     6.3  &  0.1364  &  1.6  & \phm{$-$}2.2    &   \\
              &                &               & 9.2657 &     I     & 2 &    180.8  &     0.3  &  0.1362  &  0.2  & \phm{$-$}1.7    &   \\
15471$-$5107  & WSI  83 Ba,Bb  & HD  140662B   & 8.5478 &     y     & 2 &     51.5  &     5.2  &  0.0763  &  3.4  & \phm{$-$}0.5~~: & * \\
              &                &               & 9.2603 &     y     & 2 &     48.0  &     0.6  &  0.0725  &  3.4  & \phm{$-$}0.4~~: &   \\
16057$-$3252  & WSI  84 Ba,Bb  & HIP  78842    & 8.5479 &     y     & 2 &    124.4  &     1.8  &  0.1281  &  2.5  & \phm{$-$}0.1~~: & * \\
              &                &               & 9.2630 & H$\alpha$ & 2 &    103.9  &     1.3  &  0.1202  &  0.1  & \phm{$-$}0.0    &   \\
              &                &               & 9.2630 &     y     & 2 &    103.1  &     1.6  &  0.1178  &  0.0  & \phm{$-$}0.0    &   \\
16090$-$0939  & WSI  85        & HIP  79122    & 8.5370 &     y     & 4 &    134.4  &     1.7  &  0.1423  &  1.3  & \phm{$-$}3.8    & * \\
              &                &               & 9.2631 &     y     & 2 &    135.7  &     0.6  &  0.1294  &  3.0  & \phm{$-$}3.7    &   \\
              &                &               & 9.2631 & H$\alpha$ & 2 &    134.7  &     0.8  &  0.1254  &  0.8  & \phm{$-$}3.2    &   \\
16253$-$4909  & TOK  50 Aa,Ab  & HIP  80448    & 9.2630 &     y     & 2 &    193.0  &     6.4  &  0.2286  &  1.3  & \phm{$-$}3.6    & * \\
16385$-$5728  & TOK  51 Aa,Ab  & HIP  81478    & 9.2630 &     y     & 2 &     62.2  &     0.6  &  0.2675  &  1.0  & \phm{$-$}4.0    & * \\
16534$-$2025  & WSI  86        & HIP  82621    & 8.5370 &     y     & 2 &    167.0  &     5.0  &  0.3593  &  2.9  & \phm{$-$}5.4    &   \\
17066$+$0039  & TOK  52 Ba,Bb  & HIP  83716    & 9.2659 &     y     & 3 &    181.6  &     5.2  &  0.0993  &  1.5  &       $-$0.1    & * \\
17157$-$0949  & TOK  53 Ba,Bb  & HIP  84430    & 9.2658 &     y     & 2 &    140.9  &     0.1  &  0.0328  &  0.2  & \phm{$-$}0.2    & * \\
              &                &               & 9.2658 & H$\alpha$ & 2 &    130.9  &     0.8  &  0.0365  &  0.0  & \phm{$-$}0.2    &   \\
17248$-$5913  & WSI  87 AD     & HIP  85216    & 8.5399 &     y     & 2 &    270.1  &     0.2  &  0.2673  &  2.4  & \phm{$-$}1.0    & * \\
              &                &               & 8.5399 & H$\alpha$ & 2 &    269.6  &     3.0  &  0.2675  &  3.1  & \phm{$-$}1.3~~: &   \\
              &                &               & 8.5399 &     y     & 1 &    270.0  &     0.0  &  0.2620  &  0.0  & \phm{$-$}0.5~~: &   \\
              &                &               & 9.2631 &     y     & 2 &    270.9  &     0.8  &  0.2623  &  1.0  & \phm{$-$}0.4    &   \\
              &                &               & 9.2631 & H$\alpha$ & 2 &    270.6  &     1.8  &  0.2626  &  1.8  & \phm{$-$}0.5    &   \\
              &                &               & 9.2656 & H$\alpha$ & 2 &    271.0  &     0.4  &  0.2625  &  1.2  & \phm{$-$}0.5    &   \\
              &                &               & 9.2656 &     y     & 2 &    271.4  &     0.4  &  0.2620  &  2.2  & \phm{$-$}0.5    &   \\
17390$+$0240  & WSI  88        & HIP  86374    & 8.5372 &     y     & 2 &      2.9  &     1.0  &  0.1771  &  0.7  & \phm{$-$}2.8    & * \\
              &                &               & 9.2605 &     y     & 2 &      3.8  &     0.5  &  0.1797  &  0.3  & \phm{$-$}2.6    &   \\
              &                &               & 9.2605 & H$\alpha$ & 2 &      4.9  &     0.7  &  0.1797  &  0.4  & \phm{$-$}2.5    &   \\
17535$-$0355  & TOK  54        & V2610 Oph     & 9.2605 &     y     & 3 &    149.1  &     1.3  &  0.1138  &  0.8  & \phm{$-$}0.9~~: & * \\
              &                &               & 9.2659 &     y     & 2 &    147.6  &     1.0  &  0.1074  &  3.6  & \phm{$-$}1.0~~: &   \\
17575$-$5740  & TOK  55 Ba,Bb  & HIP  87914    & 9.2632 &     y     & 2 &    179.3  &     6.2  &  0.1156  &  5.8  & \phm{$-$}0.4    & * \\
              &                &               & 9.2632 &     y     & 2 &    180.5  &     0.5  &  0.1102  &  0.9  & \phm{$-$}0.8    &   \\
              &                &               & 9.2656 &     y     & 2 &    181.1  &     0.7  &  0.1145  &  4.6  & \phm{$-$}0.2    &   \\
18024$+$2050  & TOK  56        & HIP  88331    & 9.2607 & H$\alpha$ & 2 &    301.9  &     1.2  &  0.0532  &  0.2  & \phm{$-$}1.1    & * \\
18112$-$1951  & TOK  57 Aa,Ab  & HIP  89114    & 8.7694 &     y     & 2 &    122.0  &     3.5  &  0.0396  &  4.3  & \phm{$-$}2.3    & * \\
              &                &               & 8.7694 & H$\alpha$ & 2 &    105.5  &     2.8  &  0.0492  &  3.5  & \phm{$-$}3.4    &   \\
              &                &               & 9.2634 &     y     & 2 &     27.9  &     7.1  &  0.0600  &  0.6  & \phm{$-$}3.7    &   \\
18126$-$7340  & TOK  58 Aa,Ab  & HIP  89234    & 8.7724 & H$\alpha$ & 3 &    108.3  &     1.6  &  0.3247  &  5.4  & \phm{$-$}3.8    & * \\
              &                &               & 9.2632 & H$\alpha$ & 2 &    109.9  &     0.7  &  0.3149  &  2.9  & \phm{$-$}3.6    &   \\
              &                &               & 9.2632 & H$\alpha$ & 2 &    110.3  &     0.8  &  0.3126  &  0.7  & \phm{$-$}3.6    &   \\
18152$-$2044  & TOK  59        & HIP  89439    & 8.6055 &     y     & 2 &     76.4  &     1.5  &  1.2709  &  1.4  & \phm{$-$}5.2~~* & * \\
              &                &               & 8.6055 & H$\alpha$ & 4 &     76.3  &     2.6  &  1.2728  &  2.5  & \phm{$-$}5.4~~* &   \\
18177$-$1940  & WSI  89 Ba,Bb  & HIP  89647    & 8.5425 &     y     & 2 &      5.3  &     0.1  &  0.0550  &  5.6  & \phm{$-$}0.8~~: & * \\
              &                &               & 8.6054 &     y     & 2 &      0.7  &     0.4  &  0.0590  &  1.1  & \phm{$-$}1.0    &   \\
              &                &               & 8.6054 & H$\alpha$ & 2 &      8.7  &     0.7  &  0.0541  &  3.3  & \phm{$-$}1.0    &   \\
18237$+$2146  & TOK  60 Aa,Ab  & HIP  90139    & 9.2607 & H$\alpha$ & 2 &    280.1  &     0.4  &  0.0420  &  0.3  & \phm{$-$}1.6    & * \\
18389$-$2103  & WSI  90        & HIP  91438    & 8.5372 &     y     & 2 &    241.9  &    11.7  &  0.0481  &  3.3  & \phm{$-$}2.5    & * \\
              &                &               & 8.5425 &     y     & 1 &    262.4  &     0.4  &  0.0366  &  0.4  & \phm{$-$}2.2~~: &   \\
              &                &               & 8.5425 & H$\alpha$ & 1 &    256.9  &     0.4  &  0.0463  &  0.4  & \phm{$-$}2.1    &   \\
              &                &               & 9.2634 &     y     & 2 &    137.7  &     0.4  &  0.0376  &  0.1  & \phm{$-$}2.4    &   \\
              &                &               & 9.2634 & H$\alpha$ & 1 &    151.5  &     0.1  &  0.0411  &  0.1  & \phm{$-$}2.0    &   \\
19258$-$3006  & WSI  91 Ba,Bb  & HD  182433B   & 8.5400 &     y     & 2 &    104.8  &     6.0  &  0.0443  &  1.3  & \phm{$-$}0.9    & * \\
              &                &               & 8.6055 &     y     & 2 &     92.0  &     3.0  &  0.0441  &  2.7  & \phm{$-$}0.9~~: &   \\
              &                &               & 8.6055 & H$\alpha$ & 2 &     88.2  &     6.7  &  0.0463  &  5.6  & \phm{$-$}0.8~~: &   \\
              &                &               & 8.7695 &     y     & 2 &     95.1  &     3.4  &  0.0459  &  0.1  & \phm{$-$}1.0    &   \\
              &                &               & 9.2633 &     y     & 3 &     90.1  &     2.0  &  0.0455  &  3.3  & \phm{$-$}0.7    &   \\
20401$-$2852  & SEE 423 BC     &               & 8.5483 &     y     & 3 &    105.2  &     1.9  &  0.1993  &  2.3  & \phm{$-$}0.2~~: & * \\
              &                &               & 8.6055 &     y     & 2 &    105.2  &     3.8  &  0.2035  &  2.2  & \phm{$-$}0.3~~: &   \\
22438$+$0353  & WSI  92        & HIP 112229    & 8.5376 &     y     & 2 &    118.8  &     2.9  &  1.0154  &  3.1  & \phm{$-$}4.7    & * \\
22474$+$1749  & WSI  93        & HIP 112506    & 8.5377 &     y     & 1 &    110.0  &     1.7  &  0.3054  &  1.7  & \phm{$-$}3.2    & * \\
              &                &               & 8.5377 & H$\alpha$ & 3 &    111.2  &     1.7  &  0.3053  &  0.6  & \phm{$-$}2.9~~: &   \\
              &                &               & 8.7670 &     y     & 2 &    111.5  &     0.8  &  0.3049  &  0.8  & \phm{$-$}3.1    &   \\
              &                &               & 8.7670 & H$\alpha$ & 2 &    111.3  &     0.6  &  0.3039  &  1.3  & \phm{$-$}2.7    &   \\
23444$-$7029  & WSI  94        & HIP 117105    & 8.5379 &     y     & 2 &     90.8  &     2.8  &  0.0463  &  1.2  & \phm{$-$}2.0    & * \\
23452$+$0814  & WSI  95 Aa,Ab  & HIP 117164    & 8.5378 &     y     & 2 &    194.0  &     5.2  &  1.1558  &  5.6  & \phm{$-$}4.8~~* & * \\
\enddata                                                                                                                                  
\end{deluxetable}                                                                                                                         

\clearpage
\begin{table}
\caption{Notes to individual systems (fragment)}
\label{tab:notes}
\medskip
\footnotesize
\begin{tabular}{l l p{5in}}
WDS (2000) & Discoverer    & Note \\
           & Designation   &      \\
           & or other name &      \\
\vspace{-5pt}\\
\tableline
\vspace{-5pt}\\
00028$+$0208 & BU   281 AB    & C at 44$''$ is optical \\
00059$+$1805 & STF 3060 AB    & CPM  in WDS \\
00059$-$3020 & RST 5180 AB    & dy=1.2, dm(WDS)=0.2. The status of C at 5$''$ is unknown.   \\
00063$-$4905 & HDO  180       & Companion at 4$''$, outside field. A has a planetary companion. \\
00098$-$3347 & SEE    3       & Bad orbit. \\
00121$-$5832 & RST 4739       & dm(Blanco) too large, WDS: dm=0.17mag \\
00174$+$0853 & A   1803 AB    & WDS lists 3 more companions, but only C at 3.4$''$ is physical (MSC). Two orbits for AB. C component itself is unresolved. \\
00271$-$0753 & A    431       & dy=0.6, dm(WDS)=0.11. \\
\tableline
\end{tabular}
\end{table}

\end{document}